\documentclass[fleqn,usenatbib]{mnras}
\usepackage{lastpage}
\usepackage[T1]{fontenc}
\usepackage{lastpage}
\usepackage{graphicx}
\usepackage{verbatim}
\usepackage{txfonts}
\usepackage{float}
\usepackage{color}
\usepackage{amsbsy}
\usepackage{gensymb}
\usepackage{textcomp}
\usepackage{dcolumn}
\usepackage{natbib}
\usepackage{multirow}

\usepackage{xspace}
\usepackage{hyperref}
\usepackage{caption}

\usepackage{tabularx}
\usepackage{subfig}
\usepackage{amssymb}	% Extra maths symbols
\usepackage{epsfig}
\usepackage{upgreek}
\usepackage{mathptmx}
\usepackage[utf8]{inputenc}
\usepackage{enumerate}
\usepackage{enumitem}   
\usepackage{longtable,lscape}

\usepackage{grffile}
\usepackage{array,float}
\usepackage{lscape}
\usepackage{pdflscape}
\usepackage{afterpage}
\usepackage{siunitx}
\usepackage{xargs}

\usepackage[autostyle]{csquotes}

\usepackage{setspace}

\usepackage[colorinlistoftodos]{todonotes}
\usepackage{adjustbox}
\usepackage{booktabs}
\usepackage{soul}
\usepackage{ulem}
\usepackage{siunitx}
\usepackage{lscape}
\newcolumntype{d}[1]{D{.}{\cdot}{#1}}
\definecolor{mygray}{gray}{0.6}
\newcolumntype{.}{D{.}{.}{-1}}

\bibpunct[; ]{(}{)}{,}{a}{}{;}

\newcommand{\lsun}{L$_\odot$}

\newcommand{\msun}{M$_\odot$}

\newcommand{\vlsr}{V$_{\rm{LSR}}$}

\newcommand{\mum}{$\mu$m}

\newcommand{\kms}{km\,s$^{-1}$\xspace}

\newcommand{\hii}{H\textsc{ii}}
\newcommand{\co}[2]{CO\,(#1--#2)}

\newcommand{\SC}{solar circle}
\newcommand{\higal}{Hi-GAL}
\newcommand{\oghres}{OGHReS}

\newcommand{\withVLSR}{{\color{black}2\,781}}

\newcommand{\allnonDetections}{{\color{black} 422}}

\title[OGHReS-HiGAL Outer Galaxy Clumps]{OGHReS: Star formation in the Outer Galaxy II ($\ell = 180\degr$-$280\degr$)\thanks{The full version of Tables\,\ref{tbl:velocity_allocation} and \ref{tbl:derived_clump_para} are only available in electronic form at the CDS via anonymous ftp to cdsarc.u-strasbg.fr (130.79.125.5) or via http://cdsweb.u-strasbg.fr/cgi-bin/qcat?J/MNRAS/.}}
\author[J.\,S.\,Urquhart et al.]{J.\,S.\,Urquhart,$^{1}$\thanks{E-mail: j.s.urquhart@kent.ac.uk}
C.\,K\"onig,$^{2}$ D.\,Colombo,$^{3,2}$ A.\,Karska,$^{2,3,4}$  A.\,Giannetti,$^{5}$ T.\,J.\,T.\,Moore,$^{6}$ \newauthor 
A.\,Y.\,Yang,$^{7,8}$\thanks{E-mail: yangay@nao.cas.cn}  F.\,Wyrowski,$^{2}$ Y.\,Sun,$^{9}$ Z.\,Jiang,$^{9}$ K.\,R.\,Neralwar,$^{2}$ D.\,Eden,$^{10,11}$ I.\,Grozdanova$^{1}$ \newauthor S.\,Neupane,$^{2}$ M.\,Figueira,$^{2,12}$ E.\,Dann,$^{13,2}$,   V.\,S.\,Veena,$^{2}$ W.-J. Kim,$^{13}$ S.\,Leurini,$^{14}$ J.\,Brand,$^{5}$ \newauthor M.-Y.\,Lee,$^{15}$ %\newauthor J.\,Brand,$^{6}$ D.\,Elia,$^{7}$ S.\,Leurini,$^{8}$ M.\,Figueira,$^{9,4}$ \newauthor M.-Y.\,Lee,$^{10}$   M.\,Dumke,$^{11,2}$
\\
$^{1}$ Centre for Astrophysics and Planetary Science, University of Kent, Canterbury, CT2\,7NH, UK \\
$^{2}$ Max-Planck-Institut f\"ur Radioastronomie (MPIfR), Auf dem H\"ugel 69, 53121 Bonn, Germany\\
$^{3}$ Argelander-Institut f\"ur Astronomie, Universit\"at Bonn, Auf dem H\"ugel 71, 53121 Bonn, Germany \\
$^{4}$ Institute of Astronomy, Faculty of Physics, Astronomy and Informatics, Nicolaus
Copernicus University, Grudzi\k{a}dzka 5, 87-100 Toruń, Poland\\
$^{5}$ INAF - Istituto di Radioastronomia, Via P. Gobetti 101, I-40129 Bologna, Italy\\
$^{6}$ Astrophysics Research Institute, Liverpool John Moores University, Liverpool Science Park, 146 Brownlow Hill, Liverpool, L3\,5RF, UK\\
$^{7}$ National Astronomical Observatories, Chinese Academy of Sciences, Beijing 100101, China \\
$^{8}$ Key Laboratory of Radio Astronomy and Technology, Chinese Academy of Sciences, A20 Datun Road, Chaoyang District, Beijing, 100101, P. R. China\\ 
$^{9}$ Purple Mountain Observatory, Chinese Academy of Sciences, Nanjing 210008, China\\ 
$^{10}$ Department of Physics, University of Bath, Claverton Down, Bath BA2\,7AY, UK\\
$^{11}$ Armagh Observatory and Planetarium, College Hill, Armagh BT61\,9DB, UK\\
$^{12}$National Centre for Nuclear Research, Pasteura 7, 02-093, Warszawa, Poland\\
$^{13}$ Physikalisches Institut, Universität zu K\"oln, Z\"ulpicher Str. 77, D-50937 K\"oln, Germany\\
$^{14}$ INAF - Osservatorio Astronomico di Cagliari, Via della Scienza 5, I-09047 Selargius (CA), Italy\\
$^{15}$ Korea Astronomy and Space Science Institute, 776 Daedeok-daero, 34055 Daejeon, Republic of Korea 
}

\date{Accepted XXX. Received YYY; in original form ZZZ}

\pubyear{2021}

\begin{document}
\label{firstpage}
\pagerange{\pageref{firstpage}--\pageref{lastpage}}
\maketitle

\begin{abstract}

%The Outer Galaxy High-Resolution Survey (OGHReS) covers 100\,sq degrees of the Galaxy ($180\degr < \ell < 280\degr$) using the CO (2-1) transition. We use spectra to refine the velocities, distances, and physical properties of the 6\,706 \higal\ clumps located in the OGHReS region. In a previous paper, we analysed the 3\,584 clumps between $\ell = 250\degr $ to $280\degr$ and in this paper we use the CO spectra to refine the physical parameters of the 3\,122 clumps ($180\degr < \ell < 250\degr$).  We determine reliable velocities for \withVLSR\ clumps and find good agreement with the previously assigned velocities ($\sim$80\,per\,cent within 5\,\kms). We update velocities for 288 clumps, and provide velocities to an additional 411 clumps. Combining these with the previous results, we have velocities and physical properties for 6\,193 clumps (92.3\,per\,cent). The \allnonDetections\ non-detections are low surface density clumps or likely contamination by evolved stars and galaxies. Key findings: i) improved correlation between clumps and spiral arm loci, and the discovery of clumps beyond the outer arm, supports the existence of a new spiral structure; ii) decreasing trend in the $L/M$-ratio consistent with less high-mass star formation in the outer Galaxy; iii) increase in the star formation fraction (SFF) in the outer Galaxy, suggesting that more clumps are forming stars despite their lower mass; iv) discrepancies in velocity assignments across different surveys that could affect $\sim$10\,000 clumps, especially in the fourth quadrant.

The Outer Galaxy High-Resolution Survey (OGHReS) covers 100\,square degrees ($180\degr < \ell < 280\degr$) in the (2--1) transitions of three CO-isotopologues. We use the spectra to refine the velocities and physical properties to 6\,706 \higal\ clumps located in the OGHReS region. In a previous paper, we analysed 3\,584 clumps between $\ell = 250\degr $ and $280\degr$. Here, we cover a further 3\,122 clumps ($180\degr < \ell < 250\degr$) and determine reliable velocities for \withVLSR\ of these, finding good agreement with the previously assigned velocities ($\sim$80\,per\,cent within 5\,\kms). We update velocities for 288 clumps and provide new values for an additional 411. Combining these with the previous results, we have velocities and physical properties for 6\,193 clumps (92.3\,per\,cent). The \allnonDetections\ non-detections are low surface density clumps or likely contamination by evolved stars and galaxies. Key findings: i) improved correlation between clumps and spiral arm loci, and the discovery of clumps beyond the outer arm supports the existence of a new spiral structure; ii) decreasing trend in the $L/M$-ratio consistent with less high-mass star formation in the outer Galaxy; iii) increase in the star formation fraction (SFF) in the outer Galaxy, suggesting that more clumps are forming stars despite their lower mass; iv) discrepancies in velocity assignments across different surveys that could affect $\sim$10\,000 clumps, especially in the fourth quadrant.

% word count is 236.

\end{abstract}


% --------------------- Section 2 -------------

\section{Introduction}

\begin{figure*}
    \centering
    \includegraphics[width = 0.7\textwidth, trim=20 0 20 0]{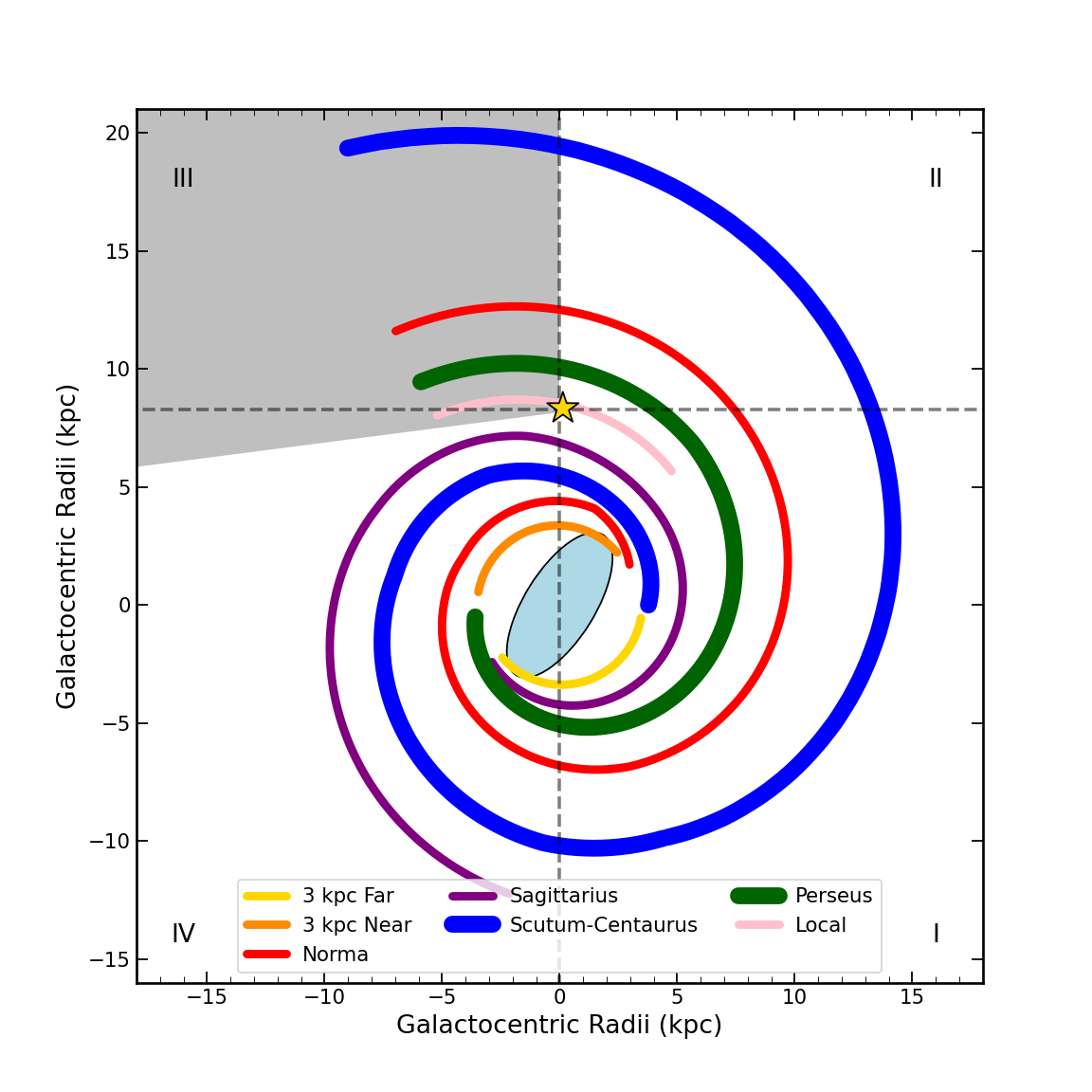}
    \caption{Schematic showing the loci of the spiral arms according to the model of \citet{reid2019}. We use the recently updated Scutum-Centaurus loci by \citet{sun2015, sun2024} and  include an additional bisymmetric pair of arm segments added to represent the 3\,kpc arms. The grey shaded region indicates the coverage of the OGHReS survey while the light grey shaded region shows the longitude coverage of the region studied in this work (see text for details). The star shows the position of the Sun and the Roman numerals identify the Galactic quadrants. The light blue oval feature located towards the centre of the diagram shows the position and orientation of the Galactic bar.}
    \label{fig:coverage_map}
\end{figure*}

The Outer Galaxy High Resolution Survey (OGHReS; {\color{blue}K\"onig et al. in prep.}) is a systematic high-resolution (i.e., $\theta_{\rm\, FWHM}\approx 30$\,arcsecs) survey of the southern outer Galactic plane in $^{12}$\co{2}{1}, $^{13}$\co{2}{1}, and C$^{18}$O\,(2-1), mapping 100\,deg$^2$ between $180\degree < \ell < 280\degree$ and 1\degr\ in $b$ using the Atacama Pathfinder Experiment 12-m submillimetre telescope (APEX;  \citealt{gusten2006}). This survey provides a $\sim$16-fold increase in angular resolution compared to the only other unbiased CO survey of this region (i.e., by \citealt{dame2001}). It allows for a detailed census of thousands of molecular clouds and filaments (\citealt{colombo2021}) and the identification of significant spurs and/or connecting bridges present in this part of the Galaxy. 

Observations of nearby galaxies have revealed vastly different star formation rates, with lower rates in irregular and dwarf galaxies than in spiral galaxies (\citealt{kennicutt2012}). Star formation rates are also found to vary significantly within galaxies due to different environmental factors  (e.g., density, metallicity, location).  Environmental conditions change enormously over the disc of the Milky Way, from the high UV-radiation flux, density, turbulence, and cosmic-ray flux found in the Galactic centre to the low-density and low metallicity environment found in the outer Galaxy (e.g., \citealt{heyer2015}, \citealt{rudolph1997}, \citealt{bloemen1984}). The large range of physical conditions and the high-spatial resolution that is available make the Milky Way the ideal place to determine the role played by environmental effects in the formation and evolution of molecular clouds, and how these, in turn, affect star formation. The main objectives of this survey are to map the distribution of molecular gas in the outer Galaxy and to investigate star formation under very different conditions from those found in the inner Galaxy and Galactic centre region.

In a recent paper (\citealt{urquhart2024}; hereafter Paper\,I), we used data acquired for the $250\degr < \ell < 280\degr$ and $-2\degr < b < -1\degr$ region to verify and, where necessary, update the velocities of some $\sim$3\,500 HiGAL sources (\citealt{elia2021}). We identified $\sim 600$ clumps where the velocity assigned from lower resolution molecular lines was incorrect and assigned velocities to a further $\sim 700$ where a velocity was not previously available. The improved velocity determination resulted in a significant increase in the proportion of sources in this region for which velocities and reliable physical properties are now available ($\sim$96\,per\,cent).

\begin{figure*}
    \centering
    \includegraphics[width = 0.90\textwidth, trim=20 0 20 0]{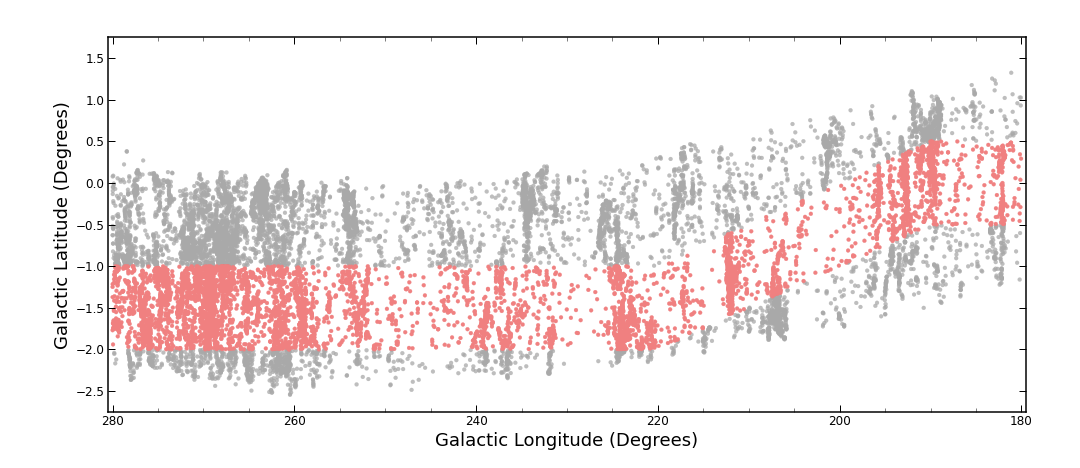}
    \caption{Distribution of \higal\ sources located within the \oghres\ longitude range. The sources covered by \oghres\ are shown as coloured (red) filled circles while those not covered in \oghres\ are shown as grey filled circles. The change in latitude was introduced by \higal\ to better follow the warp in the Galactic disc present at higher longitudes.}
    \label{fig:lb_map}
\end{figure*}

In this paper, we extend this work to the entire OGHReS survey region (i.e., $180\degr < \ell < 280\degr$).  In Figure\,\ref{fig:coverage_map}, we show the full coverage of the Galactic plane by OGHReS (grey shading) and the region covered in this work (light grey shading). We will use these data to check the velocities and distances assigned to the \higal\ clumps (\citealt{mege2021}) and refine their physical properties (\citealt{elia2021}). The structure of the paper is as follows: in Sect.\,\ref{sect:higal_cat_refining} we introduce the sample and describe the method and compare our results with \higal\ (\citealt{mege2021}); in Sect.\,\ref{sect:results} we used the updated velocities to determine distances and physical properties for the clumps; in Sect.\,\ref{sect:discussion} we look at Galactic trends in the clumps properties and discuss the implication of our results for the rest of the \higal\ catalogue (\citealt{elia2021}). In Sect.\,\ref{sect:summary} we provide a summary of the results presented here and in Paper\,I and highlight our main findings.

\section{Determining clump velocities}
\label{sect:higal_cat_refining}

The region discussed in this paper has been mapped in the continuum at far-infrared and submillimetre wavelengths by \textit{Herschel} as part of the \higal\ legacy survey (\citealt{molinari2010a}). A catalogue of $\sim$150\,000 sources covering the whole of the Galactic disc has been produced (\citealt{elia2021}). This catalogue is divided into high-reliability and low-reliability subsets, consisting of 94\,604 and 55\,619 clumps, respectively. The main difference between these two reliability types is that those identified with high-reliability have properties calculated through a spectral-energy distribution (SED) fit to the four \higal\ flux bands between 160\,\mum\ and 500\,\mum, while those identified with low-reliability have properties calculated through a fit to only three \higal\ flux bands in the same wavelength range, and consequently are considered to be less well constrained  (\citealt{elia2021}). 

The \higal\ catalogue also distinguishes between three different evolutionary stages of clumps: unbound, bound, and protostellar (see \citealt{elia2021} for more details and caveats). The distinction between bound and unbound is determined using Larson's third law ($M(r) = 460$\,\msun\,$(r/{\rm pc})^{1.9}$) as a threshold; clumps above this threshold are considered bound while clumps below it are considered unbound. The distinction between starless (bound and unbound) clumps and protostellar clumps is based on the presence or absence of a 70-\mum\ counterpart; clumps with a 70-\mum\ counterpart are classified as protostellar.

Distances are crucial for the determination of physical properties and these are primarily kinematically derived from velocity measurements taken from CO surveys (e.g., CO Heterodyne Inner Milky Way Plane Survey (CHIMPS: \citealt{rigby2016}), Structure, Excitation and Dynamics of the Inner Galactic Interstellar Medium (SEDIGISM: \citealt{schuller2017, schuller2021}), and the Galactic Ring Survey (GRS: \citealt{jackson2006})) and a Galactic rotation curve (e.g., \citealt{brand1993, russeil2017, reid2019}). At the time the \higal\ catalogue was presented, the best resolution CO surveys available in the third quadrant were the NANTEN $^{12}$CO (1--0) survey (\citealt{nagayama2011}) and the $^{12}$CO and $^{13}$CO (1--0) Exeter FCRAO Outer Galaxy survey (OGS) (\citealt{mottram2010_ogs}). The  $^{12}$CO\,(1--0) transition is easily excited at low densities and so is equally sensitive to bright diffuse clouds and dense clumps and is therefore not an ideal tracer for determining the velocity, and hence distances, to dense clumps identified by \higal.  Furthermore, the NANTEN survey has a resolution of several arcmins, which increases the probability of detecting bright emission from unrelated clouds along the line-of-sight to the clumps.

The high-resolution spectroscopic data now available from OGHReS are ideally suited to complementing the \higal\ catalogue in the 3rd quadrant. In Figure\,\ref{fig:lb_map}, we show the distribution of all \higal\ sources located within the OGHReS longitude ($\ell$) range and highlight those included in the OGHReS latitude ($b$) range. The variation in the latitude coverage of both \higal\ and OGHReS accounts for the warp in the Galactic disc in this part of the Milky Way. The OGHReS latitude range deviation from the centre of the \higal\ coverage at higher longitudes is an attempt to better probe the Perseus and Outer arms, which are known to dip below the Galactic mid-plane.

There are a total of 15\,138 \higal\ clumps located in the OGHReS longitude region (i.e., $180\degr  < \ell < 280\degr$), of which 6\,706 are also located inside the OGHReS latitude coverage. In Paper\,I, we analysed the OGHReS spectra towards 3\,584 \higal\ clumps located in the science demonstration field (i.e., $250\degr  < \ell < 280\degr$ and $-2\degr < b < -1\degr$) using the OGHReS data to update the clump velocities, distances, and physical properties. Using the kinematic information provided by the CO spectra, we were able to correct the velocities towards approximately 600 clumps and assign velocities to a further $700$ clumps where a velocity had not been previously assigned. With this more complete set of reliable velocities, we were able to determine robust physical properties and identify trends between the inner and outer parts of the Galactic disc. 

\setlength{\tabcolsep}{6pt}

\begin{table*}

\begin{center}\caption{Summary of velocity allocation analysis. In the first four columns we give the \higal\ catalogue names and Galactic coordinates and reliability, all of which have been taken from the \higal\ catalogue \citep{elia2021}. Columns 5 and 6 give the mean spectral noise for the three CO transitions and the velocity allocated to the \higal\ source from the analysis of the \oghres\ data, respectively. In Columns 7 and 8 we give the velocity allocated by the \higal\ team and the difference in velocity between the OGHReS and \higal\ velocities. In the final column we provide a note to indicate the CO transition and method used to allocate the velocity in this work (the roman numerals refer to the descriptions given in Sect.\,\ref{sec:vel}).
 }
\label{tbl:velocity_allocation}
\begin{minipage}{\linewidth}
\small
\begin{tabular}{lccc....l}
\hline \hline
  \multicolumn{1}{c}{\higal}&
  \multicolumn{1}{c}{$\ell$} & 
  \multicolumn{1}{c}{$b$} &
  \multicolumn{1}{c}{Reliability}&
  \multicolumn{1}{c}{RMS} &
  \multicolumn{1}{c}{OGHReS \vlsr} &
  \multicolumn{1}{c}{\higal\ \vlsr} &
  \multicolumn{1}{c}{$|\Delta $V$_{\rm{LSR}} |$} & 
  \multicolumn{1}{c}{Allocation} \\

  \multicolumn{1}{c}{name }&
  \multicolumn{1}{c}{(\degr)} & 
  \multicolumn{1}{c}{(\degr)} & 
  \multicolumn{1}{c}{} & 
  \multicolumn{1}{c}{(K)}&
  \multicolumn{1}{c}{(\kms)} &
  \multicolumn{1}{c}{(\kms)} &
  \multicolumn{1}{c}{(\kms)} &
  \multicolumn{1}{c}{method}    \\
    
\hline

HIGALBM180.1134$-$0.2891	&	180.1134	&	$-$0.2891	&	high	&	0.16	&	0.9	&	0.6	&	0.3	&	$^{12}$CO (i)	\\
HIGALBM180.1184$+$0.0396	&	180.1184	&	$+$0.0396	&	low	&	0.13	&	-10.5	&	17.9	&	28.4	&	$^{12}$CO (i)	\\
HIGALBM180.1995$-$0.4556	&	180.1995	&	$-$0.4556	&	high	&	0.12	&	-7.8	&	-7.7	&	0.1	&	$^{12}$CO (i)	\\
HIGALBM180.3477$-$0.2586	&	180.3477	&	$-$0.2586	&	low	&	0.13	&	-2.9	&	34.0	&	36.9	&	$^{12}$CO (i)	\\
HIGALBM180.6616$+$0.0587	&	180.6616	&	$+$0.0587	&	high	&	0.17	&	6.2	&	6.5	&	0.3	&	$^{13}$CO (i)	\\
HIGALBM180.7606$-$0.1808	&	180.7606	&	$-$0.1808	&	low	&	0.21	&	3.1	&	36.5	&	33.4	&	$^{12}$CO (ii)	\\
HIGALBM180.7637$-$0.2279	&	180.7637	&	$-$0.2279	&	low	&	0.22	&	4.3	&	36.0	&	31.7	&	$^{12}$CO (i)	\\
HIGALBM180.7830$-$0.2561	&	180.7830	&	$-$0.2561	&	low	&	0.17	&	3.7	&	2.8	&	0.9	&	$^{12}$CO (ii)	\\
HIGALBM180.9142$+$0.4024	&	180.9142	&	$+$0.4024	&	high	&	0.17	&	-0.9	&	-1.1	&	0.2	&	$^{13}$CO (i)	\\
HIGALBM180.9144$+$0.3910	&	180.9144	&	$+$0.3910	&	high	&	0.17	&	-0.7	&	-0.7	&	0.0	&	$^{13}$CO (i)	\\

\hline\\
\end{tabular}\\
Notes: Only a small portion of the data is provided here. The full table is available in electronic form at the CDS via anonymous ftp to cdsarc.u-strasbg.fr (130.79.125.5) or via http://cdsweb.u-strasbg.fr/cgi-bin/qcat?J/MNRAS/.
\end{minipage}

\end{center}
\end{table*}
\setlength{\tabcolsep}{6pt}

\begin{figure*}
	\centering

        \includegraphics[width=0.49\textwidth]{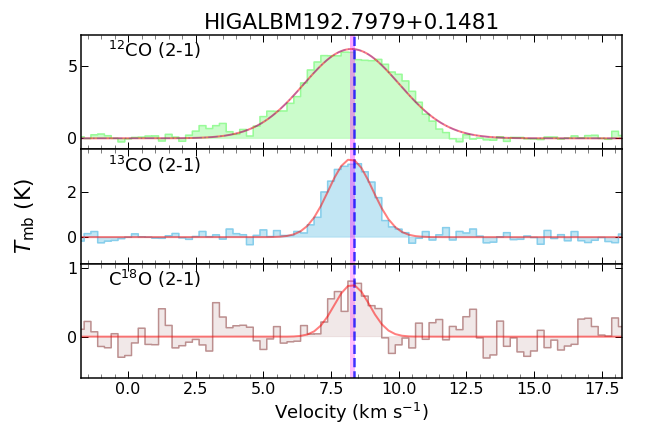}
        \includegraphics[width=0.49\textwidth]{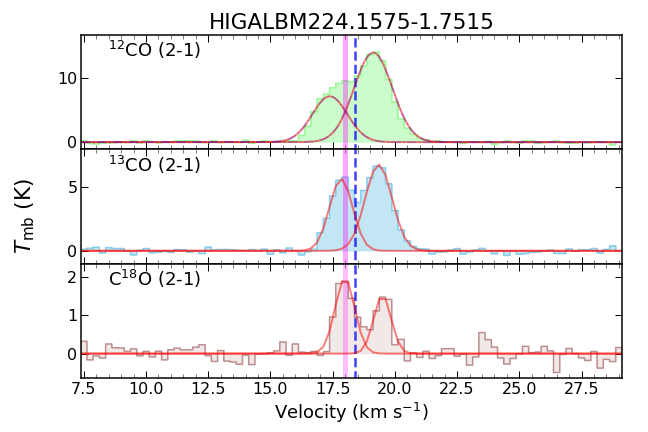}\\
        \includegraphics[width=0.49\textwidth]{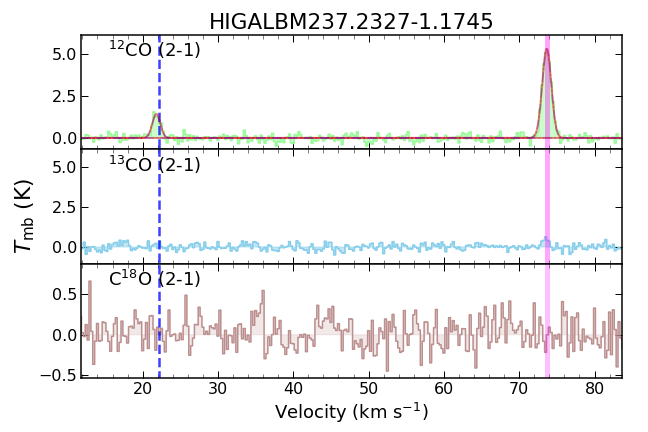}
        \includegraphics[width=0.49\textwidth]{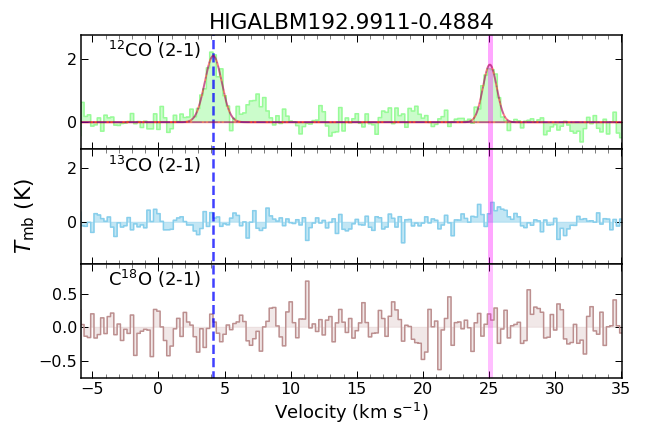}\\
    	\caption{CO\,(2--1) line emission toward selected \higal\ clumps. The coloured spectra show the data while the red curve shows the Gaussian fits. The blue vertical dashed line shows the velocity allocated by the \higal\ team (\citealt{mege2021,elia2021}) and the thick vertical magenta line shows the velocity allocated from the OGHReS data in this paper. These have been selected to illustrate the method used to allocate velocities to the \higal\ clumps (see text for details): the upper left panel shows a single component where the allocation is unambiguous; the upper right panel shows an example where components are blended, in these cases we selected the brightest component of C$^{18}$O emission; in the lower left panel we show an example where multiple components are detected widely separated in velocity, in these cases and where one component is significantly brighter we selected the brightest as the most likely component; in the lower right panel we show an example where two approximately equally bright components are detected, in these cases we compare the morphologies of the components from integrated CO maps to determine the most likely component.}
		\label{fig:velocity_examples}
	
\end{figure*}

In this paper, we analyse the OGHReS CO spectra towards the 3\,122 \higal\ clumps located in the remaining part of the OGHReS region (i.e., $180\degr  < \ell < 250\degr$). Of these, 1\,688 are flagged as being highly reliable, with the remaining 1\,434 being given the low-reliability flag in the Hi-GAL catalogue. This sample includes 662 protostellar, 1\,128 bound, and 1\,332 unbound clumps. However, we note that the distinction between bound and unbound is somewhat dependent on the distance, which is itself dependent on the velocity, and so if the velocity of a clump is updated, it may affect its bound/unbound classification (see Section\,\ref{sect:physical_properties}).

\subsection{Velocity determination}
\label{sec:vel}

 We extracted $^{12}$CO\,(2--1),  $^{13}$CO\,(2--1) and C$^{18}$O\,(2--1) spectra towards all 3\,122 clumps covered by both \higal\ and OGHReS. To increase the signal-to-noise ratio, we averaged the emission over the beam (i.e., the closest 9 pixels, given the pixel size of $9.5\times 9.5$\,arcsec$^2$). The spectra cover a velocity range from $-$50\,\kms to 150\,\kms and have a velocity resolution of 0.25\,\kms.
 Subsequently, a fifth-order polynomial function is fitted to the emission-free channels to correct for variations in the spectral baseline.\footnote{The high order polynomial fit was needed to accommodate broad undulations present in a baseline of the 200\,\kms velocity range searched for emission.  }

 The fitting and steps used to assign velocities to each clump are described in Paper\,I, and so only a brief overview is presented here, and we refer the reader to that paper for more details. The noise is determined from the standard deviation of the emission-free channels. Peaks above the 4-$\sigma$ noise threshold ($\sigma \approx 0.2$\,K\,channel$^{-1}$ for all transitions)  are identified using the \texttt{find\_peaks} function and fitted simultaneously using the \texttt{curve\_fit} function to the data with a Gaussian function.\footnote{Both functions are part of the Python \texttt{SciPy} package (\citealt{scipy}).} All fits are visually checked to ensure that their results are reliable and optimised where necessary (primarily needed for blended emission components). We require that emission covers at least three contiguous 0.25-\kms\ channels over the 3-$\sigma$ threshold to be considered a detection.

In total, we have identified 4\,740 $^{12}$CO spectral components towards 2\,803 \higal\ clumps and 2\,587 $^{13}$CO spectral components towards 2\,158 \higal\ clumps. In addition to the $^{12}$CO and $^{13}$CO, we analysed the C$^{18}$O spectra for the whole OGHReS-\higal\ sample of 6\,706 clumps detecting emission towards 913 sources, 514 of which are located in the region focused on in this paper. There are 319 clumps towards which no emission has been detected. 

\begin{figure*}
\centering
        \includegraphics[width=0.45\textwidth]{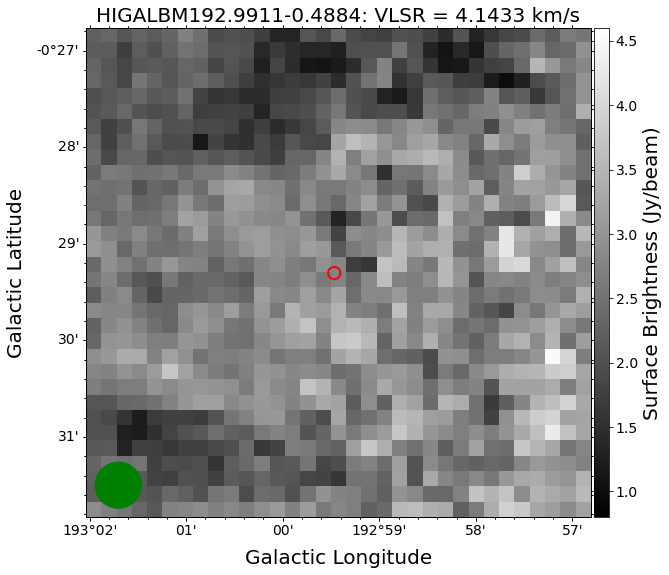}
        \includegraphics[width=0.45\textwidth]{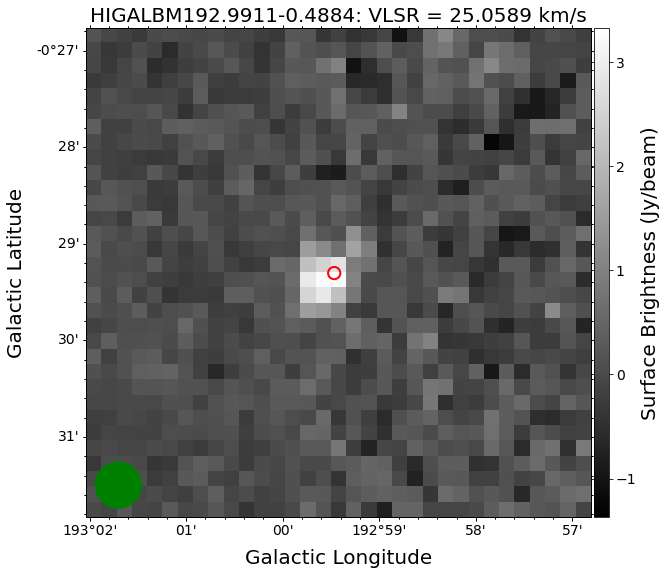}
    
   	\caption{Integrated emission maps for HIGALBM192.9911$-$0.4884 for the two approximately equal $^{12}$CO components detected at different velocities that make assignment of a velocity from the spectra alone unreliable (see lower right panel of Fig.\,\ref{fig:velocity_examples}). The beam size is shown as a green circle in the lower left corner and the position of the \higal\ source is indicated by the red circle. The velocity of the peak component about which the emission is integrated is given above each map. The coincidence of the \higal\ sources with compact molecular gas at $\sim$25\,\kms\ allows us to allocate this velocity to the source with a high degree of confidence.   
   	}
	\label{fig:integrated_co_maps}
	
\end{figure*}

When assigning velocities to clumps we are making the simplifying assumption that the \higal\ clump is a distinct object at a single velocity and that its high density means that it is more likely to be associated with the most intense CO peak. We prioritise the C$^{18}$O and $^{13}$CO transitions when assigning velocities as these are less affected by blending and self-absorption than the more abundant $^{12}$CO transition, which is nearly always optically thick. Furthermore, their lower comparative abundance means they tend to be associated with high column density objects. Below we describe the methods used to allocate velocities in order of reliability:

\begin{enumerate} [label=(\roman*)]
    \item There are 490 clumps associated with a single C$^{18}$O peak,  1\,430 clumps associated with a single $^{13}$CO peak not detected in C$^{18}$O, and 398 clumps associated with a single $^{12}$CO peak not detected in either of the other two lines. We are therefore able to unambiguously assign a velocity to 2\,318 clumps (see top left panel of Fig.\,\ref{fig:velocity_examples} for an example).\\
    
    \item There are 510 clumps with multiple emission peaks detected. If the emission peaks are blended or in close proximity to each other (within $\pm$5\,\kms) we select the strongest peak, as the uncertainty in selecting the wrong peak is smaller than the uncertainty in the streaming motion ($\pm$7\,\kms;  \citealt{reid2014}), which dominates the uncertainty in the corresponding distance. This method is applied to all three transitions starting with C$^{18}$O and ending with the $^{12}$CO transition. This allows us to assign velocities to a further 392 clumps (see top right panel of Fig.\,\ref{fig:velocity_examples} for an example).\\

    \item In cases where the emission peaks are more widely separated in velocity we select the most intense component providing that it is at least twice the intensity of the next brightest component (c.f. \citealt{urquhart2021}). This method is again applied to all three transitions starting with C$^{18}$O and ending with the $^{12}$CO transition and has been successful in assigning velocities to 33 more clumps (see lower left panel of Fig.\,\ref{fig:velocity_examples} for an example). \\

    \item For the remaining 60 clumps where emission is seen at two or more velocities with approximately equal intensities with the given noise levels (see lower right panel of Fig.\,\ref{fig:velocity_examples} for an example) we have created $5\times 5$-arcmin integrated $^{12}$CO maps of the strongest components and selected the velocity component for which the peak of the emission most closely correlates with the position of the \higal\ clump (cf. Paper\,I). This method has been successful in allocating velocities for a further 38 clumps (see lower right panel of Fig.\,\ref{fig:velocity_examples} and Fig.\,\ref{fig:integrated_co_maps} for an example of this method). \\

\end{enumerate}

In Table\,\ref{tbl:velocity_allocation}, we give the clump names, positions, reliability and \oghres\ and \higal\ velocities, the difference in the allocated velocities and the transition and method used to allocated the velocities in this work. In total, we have determined reliable velocities for \withVLSR\  \higal\ clumps in the studied region (corresponding to $\sim$90\,per\,cent of the sample). 

Of the remaining sources 341 clumps, we have been unable to allocate a reliable velocity to 22 clumps for which multiple CO components have been detected ($\sim$2\,per\,cent) and 319 are CO non-detections ($\sim$10\,per\,cent).

\subsection{Comparison with velocities assigned to \higal\ clumps}

\begin{figure}
	\centering

        \includegraphics[width=0.45\textwidth]{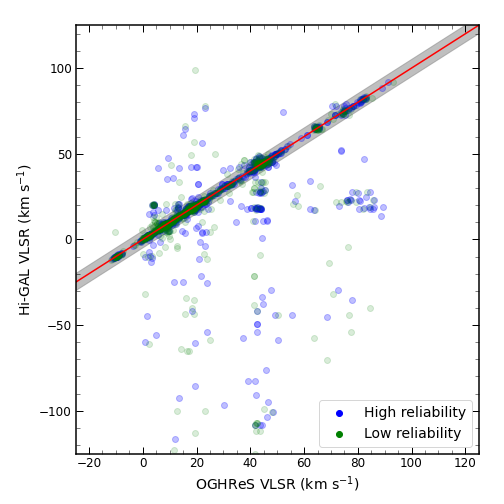}
    
    	\caption{Comparison of the velocities reported by \higal\ and the velocities determined in this paper from fits to the OGHReS spectra. The blue and green circles show the high- and low-reliability sources, respectively, and the red line shows the line of equality. The velocities are considered to be in agreement if they are within $\pm$5\,\kms; this is indicated by the grey shaded region around the line of equality.
     }
		\label{fig:velo_offsets}
	
\end{figure}

 The \higal\ catalogue provides velocities for 2\,522 clumps in OGHReS region discussed here (i.e., $180\degr < \ell < 250\degr$). Comparing these with the velocities determined from the OGHReS data we find the velocities agree in only 2\,042 cases corresponding to $\sim$80.0\,per\,cent of the sample. Of the 480 disagreements, there are 288 cases where the velocities assigned disagree by more than 5\,\kms\ (131 are classified as being low-reliability in the \higal\ catalogue, with the remaining 157 clumps being high-reliability). In Figure\,\ref{fig:velo_offsets}, we show all clumps for which velocity measurements are available in both catalogues in the $\ell$ between 180\degr\ and 250\degr. For the remaining clumps no emission is detected in OGHReS; these velocities are considered to be unreliable and have been discarded. 

In Paper\,I, we compared the velocity allocated from the \oghres\ data with those allocated by the \higal\ team (\citealt{mege2021}) and found significant differences ($> 5$\,\kms) for $\sim$ 25\,per\,cent of the sources in common. The large number of velocity disagreements was somewhat surprising, given that \citet{mege2021} used a similar methodology to assign velocities, including looking for morphological agreement in cases where multiple components were detected. In Paper\,I, we conducted a detailed analysis to understand these anomalies and concluded this was the result of the large NANTEN beam (2.7\arcmin), sparse grid spacing used (4\arcmin) and sensitivity of the $^{12}$CO (1--0) transition to low-column density molecular material. This combination of sensitivity to low-column densities and low-resolution has resulted in the detection of additional unrelated bright emission components within the beam, which made identifying the correct velocity more challenging.

\begin{figure*}
	\centering
        
        \includegraphics[width=0.90\textwidth, trim = 0cm 0.5cm 0cm 1cm, clip]{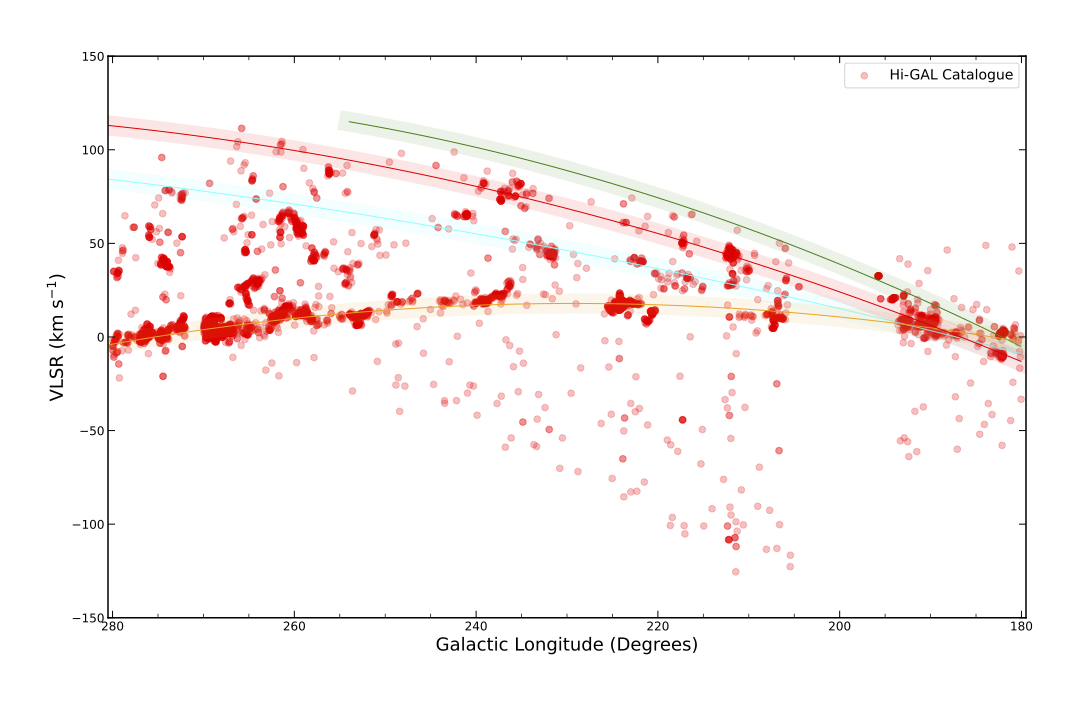}
        \includegraphics[width=0.90\textwidth, trim = 0cm 0.5cm 0cm 1cm, clip]{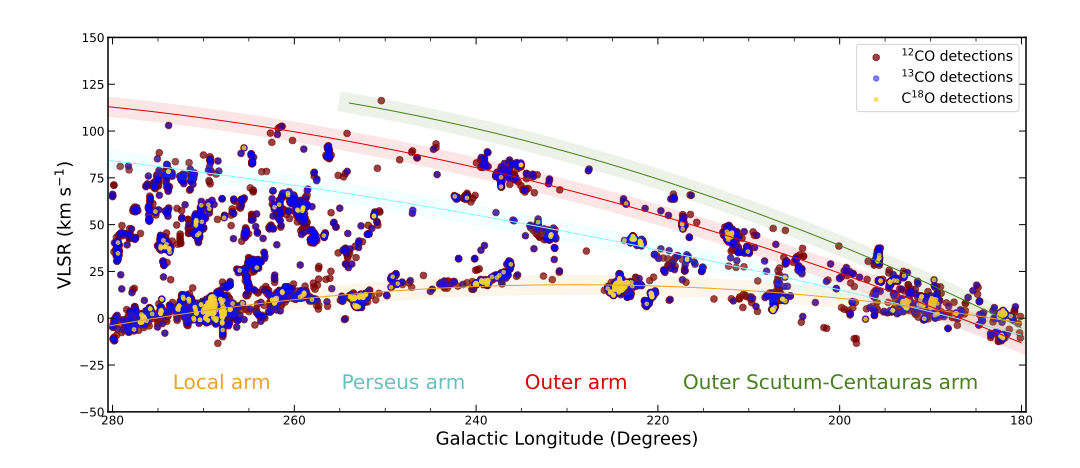}

    	\caption{Longitude-velocity distribution of \higal\ clumps located in the OGHReS survey region. The Local, Perseus and Outer arm models are shown in yellow, cyan and red, respectively, and are taken from \citet{reid2019}. The new Outer Scutum-Centaurus discovered by \citet{sun2015,sun2024} is shown in green. In the upper panel we show the distribution of velocities allocated by the \higal\ team (\citealt{mege2021}). The lower panel shows the velocities allocated from the analysis of the OGHReS data presented here and in Paper\,I. The \higal\ clumps are coloured to show the distribution of the CO transitions detected in the \oghres\ data (see legend for details).
        }
		\label{fig:lv_diagrams}
	
\end{figure*}

In the vast majority of the velocity disagreements the \higal\ velocities have been allocated using the low-resolution NANTEN $^{12}$CO (1--0) data (374 clumps, corresponding to 23\,per\,cent of clumps with velocities assigned from this transition) and so these can be explained by the difference in resolution. Of the other disagreements, $^{12}$CO and $^{13}$CO (1--0) Exeter FCRAO OGS data (\citealt{mottram2010_ogs}) was used by \higal\ to assign velocities for 43 and 35 of these respectively, corresponding to 28\,per\,cent and 5.3\,per\,cent of the clumps these transitions were used for. The high level of disagreements with the $^{12}$CO (1--0) Exeter FCRAO OGS data is hard to explain but this transition was only used to assign velocities to a small fraction of clumps (154). The much higher level agreement with the $^{13}$CO (1--0) Exeter FCRAO OGS data ($\sim$95\,per\,cent) is somewhat more reassuring.

There are 600 clumps in the $\ell$ range between 180\degr\ and 250\degr\ that \higal\ was unable to allocate a velocity for, including the longitude region between 195\degr\ and 205\degr\ where no CO line survey data was available. Using the OGHReS data we are able to allocate velocities to 424 clumps with a further 170 found to be non-detections. For the remaining six clumps we were unable to reliably allocate a velocity from the multiple components detected in the CO data.

OGHReS, therefore, has significantly improved angular and spectral resolutions compared to the NANTEN survey. In addition, OGHReS uses a higher$-J$ CO transition and is more sensitive to denser gas than is traced by $^{12}$CO\,(1--0). In total, we have been able to assign reliable velocities to \withVLSR\ clumps, which corresponds to 92\,per\,cent of the clumps in this part of the survey, with no CO being detected towards the majority of the other clumps. This is a significant improvement on the proportion of clumps that previously had reliable velocity allocations ($\sim$65\,per\,cent).

\section{Characterising the \higal\ clumps}
\label{sect:results}

There are 6\,706 \higal\ clumps in the region surveyed by OGHReS. Combining the results presented here and those presented in Paper\,I, we have been able to assign velocities to 6\,193 clumps, with 422 non-detections and there are 91 clumps for which we have not been able to assign a velocity. In this section, we look at the Galactic distribution of the detections, describe how the  physical properties of the clumps are determined, and discuss the nature of the non-detections. 

\subsection{Correlation with the spiral arms}

In Figure\,\ref{fig:lv_diagrams}, we show the longitude-velocity distribution of \higal\ clumps in the \oghres\ region. In the upper panel, we show the velocities allocated by the \higal\ team, while in the lower panel, we show the velocities allocated using the \oghres\ data presented here and in Paper\,I. The \higal\ team allocated a significant number of clumps to large negative velocities; consequently, a larger velocity range is needed in the upper panel to include all of the clumps. Figure\,\ref{fig:lv_diagrams} also shows the loci for the spiral arms located in the \oghres\ region; these have been produced by fitting a polynomial to the loci given in \citet{reid2019} for the Local, Perseus and Outer arms, and the outer Scutum-Centauras (OSC) arm identified by \citet{sun2015,sun2024}. 

The plots presented in Fig.\,\ref{fig:lv_diagrams} demonstrate the significantly improved correlation of the clumps with the spiral arms obtained with the OGHReS velocities compared to those given in the \higal\ catalogue. The upper panel of this figure reveals the presence of a significant number of clumps with large negative velocities that have substantial offsets from the spiral features and are unphysical with respect to the Galactic rotation curve. These issues have been resolved in the updated velocities assigned to the clumps, as the updated distribution of clumps (lower panel of Fig.\,\ref{fig:lv_diagrams}) is devoid of negative velocity outliers and shows higher density concentrations of clumps that are correlated with the spiral arms. We also note a small but significant number of clumps that have velocities placing them beyond the Outer arm. This sample consists of approximately 10 sources located between $\ell$ of 200\degr and 250\degr and a cluster of approximately 60 clumps located at $\ell = 196$\degr. The location of these clouds would be hard to explain if not for their correlation with the outer Scutum-Centaurus arm (\citealt{sun2015, sun2024}). The correlation of these clumps with this newly discovered spiral feature provides support for its existence.

We note that, while the correlation between the clumps and spiral structure is generally good for the Local, Outer and OSC arms, it is poorer for the Perseus arm between $\ell = $ 250\degr\ and 280\degr. This region is associated with eight Galactic super-bubbles (\citealt{McClure-Griffiths2002}, \citealt{mcclure2005}) and 38 filamentary structures that have the appearance of spurs (\citealt{colombo2021}). The Perseus arm therefore appears to consist of a variety of segments, spurs and inter-arm features or waves/corrugations (cf.  \citealt{poggio2024}) that could cause the clumps to appear less correlated with the main arm structure in localised regions. This region has also been associated with large non-circular motions (see figure\,7 of \citealt{brand1993}), which makes kinematic distances in this region more uncertain.

We can speculate that the Perseus arm might have been disturbed or destroyed by the same events that caused the formation of the GSH242-03+37 super-bubble. This object (with a radius of $\sim0.5$\,kpc) is one of the largest and most energetic (expansion energy of $3\times10^{53}$ erg/s) of those structures known in the Milky Way \citep{mcclure2005}. Cavities in the ISM are not uncommon in the Milky Way (e.g. \citealt{churchwell2006}) and in nearby galaxies (see e.g. \citealt{watkins2023}), however, super-shells like GSH242-03+37 are on the upper limit of the size distribution. In NGC\,628, the expansion of the ``Phantom Void'' bubble (similar in size to GSH242-03+37) is carving a significant hole in a galaxy spiral arm (\citealt{barnes2023}), which is notable at several wavelengths.

\subsection{Kinematic Distances}

\begin{figure}
	\centering

         \includegraphics[width=0.45\textwidth]{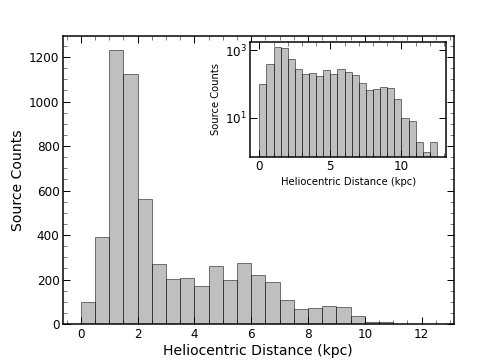}
         \includegraphics[width=0.45\textwidth]{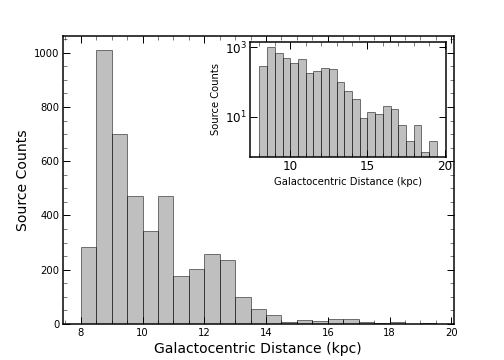}\\

    	\caption{Updated distance distributions of the HiGAL clumps. The plot inserted into the upper right corner of each plot is the same shown in the main plot but the y-axis is in log-scale to better show the distribution at large distances. The bin width used is 0.5\,kpc in both plots.}
		\label{fig:distance_velocity_hist}

\end{figure}

We combine the radial velocities assigned from the \oghres\ data with a Galactic rotation curve to determine kinematic distances and Galactocentric distances for the clumps. To ensure consistency with previous OGHReS papers (i.e., Paper\,I, \citealt{colombo2021}) and the SEDIGISM survey (\citealt{schuller2021,cabral2021})  we utilize the \citet{brand1993} rotation curve (but using $R_0 = 8.15$\,kpc and $\theta_0 =240$\,\kms; see \citealt{colombo2021} for discussion of these parameters). We have compared the distances determined with those given in the \higal\ catalogue (\citealt{elia2021}) for clumps where the velocities are closely aligned (i.e., $<$5\,\kms) and find they are in good agreement; thus, the choice of rotation model is not crucial to this work (see figure\,8 of Paper\,I).

There is no kinematic distance ambiguity for sources located outside the \SC\ (i.e., Galactocentric distances ($R_{\rm gc}) > 8.15$\,kpc), making the assigning distances somewhat more straightforward. However, distances to outer Galaxy sources are not without their idiosyncrasies. Sources close to the Galactic anti-centre are associated with very large uncertainties, as all velocities tend towards zero at this point (a similar issue affects clouds located towards the Galactic centre). Furthermore, rotation curves provide no solution for sources with negative velocities located between $\ell$ of 180\degr\ and 270\degr, and it is not possible to estimate lower limits for the distances to sources with small positive radial velocities (e.g. \vlsr\ $<5$\,\kms).

We estimate the uncertainties in the determined kinematic distances by applying a $\pm$7\,\kms\ correction to the assigned velocity to account for streaming motions and cloud-cloud velocity dispersion (i.e. random motions; see e.g., \citealt{stark1989}) and recalculating the distances using the rotation curve. This generates an upper and lower limit for the assigned distance to each clump, with the uncertainties being the difference between the limits and the distance determined from the allocated velocity. For clumps with velocities less than $< 7$\,\kms\ it is not possible to determine the lower limit to the distance uncertainty (negative velocities are unphysical in this part of the Galaxy). Comparing the upper and lower limit for clumps we are able to determine both limits for, we find them to have similar values. For sources where we are unable to estimate a lower limit for the distance we assume the positive and negative uncertainties have similar values and adopt the positive uncertainty for the source. We then convert the uncertainties into fractions of the clump distances and exclude all sources where the uncertainty in the distance is larger than a factor of 2.

Using the rotation curve and the velocities, we are able to calculate the kinematic distances to 5\,872 clumps.  However, limiting the sample to those with distances with uncertainties less than a factor of 2, reduces the number of clumps to 3\,449, which are considered to be reliable. All of the clumps with large distance uncertainties either have velocities $< 10$\,\kms\ or are located towards the Galactic anti-centre.  

\setlength{\tabcolsep}{3pt}

\begin{table*}

%select higal_name,  '\&', reliability, '\&', oghres_evol_flag, '\&', round(oghres_vlsr_final,1),'\&', FORMAT(oghres_dist_kpc,1),'\&', format(oghres_rgc_kpc,1),'\&',format(gas2dust_ratio,1), '\&', format(oghres_diam_pc,2), '\&',format(log10(oghres_mass),3), '\&',format(oghres_surface_density,3), '\&',format(log10(oghres_lum),3), '\&',format(log10(oghres_lum/oghres_mass),3),'\\\\' from oghres_higal_complete where oghres_dist_reliable is not null and gal_long > 188.9139 limit 11

\begin{center}\caption{Updated \higal\ clump parameters. The distances have been determined from the OGHReS velocities determined here and the gas-to-dust ratio ($\gamma$) from \citet{giannetti2017}. The rest of the properties have been scaled from those given in the \higal\ catalogue using the updated distances and gas-to-dust ratios. The integers given in the Evolution type column correspond to unbound = 0, starless = 1 and protostellar = 2.}
\label{tbl:derived_clump_para}
\begin{minipage}{\linewidth}
\small
\begin{tabular}{lc....c.....}
\hline \hline
  \multicolumn{1}{c}{\higal}&  
  \multicolumn{1}{c}{\higal} &
  \multicolumn{1}{c}{Evolution}&
  \multicolumn{1}{c}{\vlsr} & 
  \multicolumn{1}{c}{Distance} &
  \multicolumn{1}{c}{R$_{\rm{gc}}$}&
  \multicolumn{1}{c}{$\gamma$} &
  \multicolumn{1}{c}{Diameter} & 
  \multicolumn{1}{c}{Log[$M_{\rm{clump}}$]} & 
  \multicolumn{1}{c}{$\Sigma_{\rm gas}$} &  
  \multicolumn{1}{c}{Log[$L_{\rm{bol}}$]}	&
  \multicolumn{1}{c}{Log[$L/M$]} \\

    \multicolumn{1}{c}{name }&  
    \multicolumn{1}{c}{reliability flag}&
    \multicolumn{1}{c}{type} & 
    \multicolumn{1}{c}{(\kms)}  &
    \multicolumn{1}{c}{(kpc)} &
    \multicolumn{1}{c}{(kpc)}&
     \multicolumn{1}{c}{} &
    \multicolumn{1}{c}{(pc)}&
    \multicolumn{1}{c}{(\msun)} &
    \multicolumn{1}{c}{(g\,cm$^{-2}$)} &
    \multicolumn{1}{c}{(\lsun)} &
    \multicolumn{1}{c}{(\lsun/\msun)}   \\
    
\hline

HIGALBM188.9140-0.2402	&	low	&	1	&	16.7	&	6.7	&	14.8	&	537.3	&	0.48	&	2.163	&	0.173	&	0.655	&	-1.508	\\
HIGALBM189.2168-0.0002	&	low	&	1	&	10.9	&	3.3	&	11.4	&	270.3	&	0.60	&	1.935	&	0.064	&	0.846	&	-1.089	\\
HIGALBM189.2761-0.0707	&	low	&	1	&	13.3	&	4.3	&	12.4	&	332.2	&	0.53	&	1.624	&	0.041	&	0.724	&	-0.900	\\
HIGALBM189.2898-0.0756	&	low	&	1	&	13.1	&	4.2	&	12.3	&	323.0	&	0.41	&	2.709	&	0.822	&	0.520	&	-2.189	\\
HIGALBM189.3372-0.4986	&	high	&	1	&	16.9	&	6.3	&	14.4	&	494.9	&	0.81	&	2.779	&	0.244	&	1.024	&	-1.755	\\
HIGALBM189.3875+0.4849	&	high	&	2	&	10	&	2.8	&	10.9	&	247.0	&	0.17	&	1.890	&	0.692	&	1.651	&	-0.238	\\
HIGALBM189.4095+0.2656	&	low	&	1	&	11.1	&	3.2	&	11.4	&	267.6	&	0.39	&	1.841	&	0.119	&	0.209	&	-1.632	\\
HIGALBM189.4304-0.4217	&	high	&	1	&	10.2	&	2.9	&	11.0	&	249.5	&	0.43	&	1.917	&	0.121	&	0.268	&	-1.649	\\
HIGALBM189.4655+0.4259	&	high	&	1	&	10.9	&	3.2	&	11.3	&	263.9	&	0.50	&	1.753	&	0.060	&	1.169	&	-0.584	\\
HIGALBM189.4701+0.3893	&	low	&	1	&	11.3	&	3.3	&	11.4	&	271.4	&	0.36	&	2.024	&	0.222	&	0.439	&	-1.585	\\
HIGALBM189.4756+0.3752	&	high	&	1	&	11.2	&	3.3	&	11.4	&	269.2	&	0.25	&	1.690	&	0.214	&	0.616	&	-1.074	\\

\hline\\
\end{tabular}\\
Notes: Only a small portion of the data is provided here. The full table is available in electronic form at the CDS via anonymous ftp to cdsarc.u-strasbg.fr (130.79.125.5) or via http://cdsweb.u-strasbg.fr/cgi-bin/qcat?J/MNRAS/.
\end{minipage}

\end{center}
\end{table*}
\setlength{\tabcolsep}{6pt}

With the kinematic distances in hand it is relatively straightforward to calculate the Galactocentric distances:

\begin{equation}
R_{\rm gc}  = \sqrt{\left(d_{\rm gc}-d\,\cos \ell\right)^2 + \left( d\,\sin \ell \right)^2}
\end{equation}

\noindent where $d_{\rm gc}$ is the distance to the Galactic centre (taken as 8.15\,kpc), $d$ is the kinematic distance to the clump, and $\ell$ is the Galactic longitude of the clump. The upper and lower limit for the kinematic distances are used to estimate the uncertainties in the Galactocentric distances. Due to the relationship between the Galactocentric and kinematic distances, it is possible that clumps with large kinematic uncertainties can have significantly smaller Galactocentric uncertainties. We have determined the Galactocentric distance to a total of 5\,872 clumps, of which 5\,783 have uncertainties smaller than a factor of two.

The heliocentric and Galactocentric distance distributions of all clumps with reliable measurements are shown in the upper and lower panels of Fig.\,\ref{fig:distance_velocity_hist}, respectively. Both distributions have large numbers of clumps at low distances, indicating that the sample is dominated by nearby sources associated with the Local arm.

\subsection{Physical properties}
\label{sect:physical_properties}

\begin{figure*}
	\centering

        \includegraphics[width=0.45\textwidth]{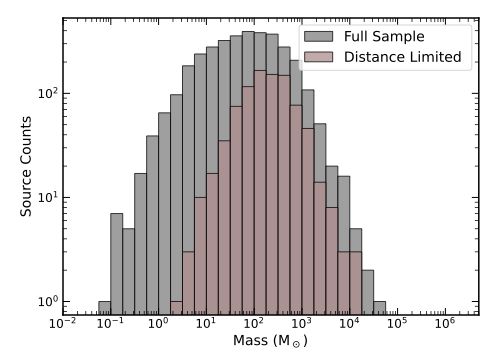}
        \includegraphics[width=0.45\textwidth]{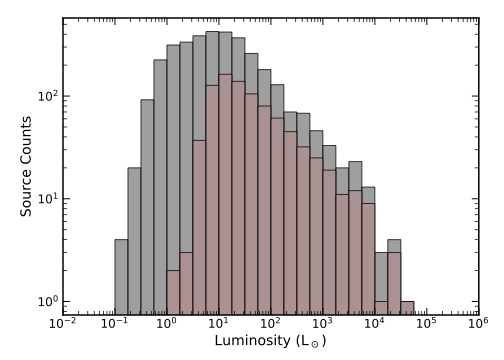}
        \includegraphics[width=0.45\textwidth]{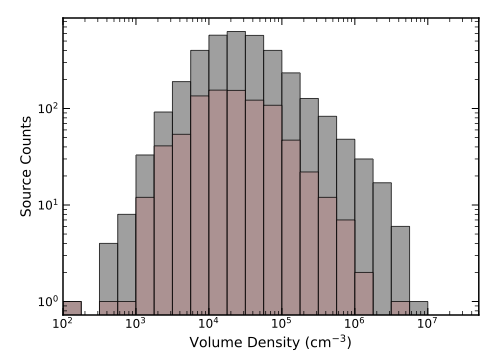}
        \includegraphics[width=0.45\textwidth]{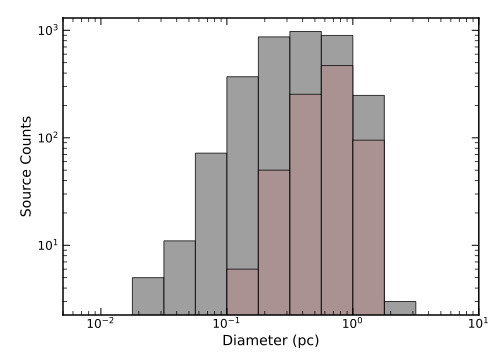}
            
    	\caption{Distributions of the four distance-dependent physical parameters. We show the full and distance-limited samples of clumps with reliable distances. The bin size in all plots is 0.25\,dex.}
		\label{fig:parameter_histo}

\end{figure*}

We use the updated distances discussed in the previous section and the variation in the gas-to-dust ratio $\gamma$ as a function of Galactocentric distance \citep{giannetti2017} to re-calculate the physical properties of the \higal\ clumps (i.e., luminosity, mass, radius, surface density and luminosity-to-mass ratio; see Paper\,I for more details). In Table\,\ref{tbl:derived_clump_para}, we provide the updated values for the \higal\ clumps and in Fig.\,\ref{fig:parameter_histo} we show the distributions of the four distance-dependent parameters for the full sample and a distance-limited sample (5-7\,kpc). The clumps have typical masses of a few hundred solar masses, diameters of $\sim$0.5\,pc and H$_2$ densities of a few times 10$^4$\,cm$^{-3}$; these are dense, high-mass clumps that are likely formation sites of future clusters.

\begin{comment}
    select higal_evol_flag, oghres_evol_flag, count(*)  from oghres_higal_complete  group by higal_evol_flag, oghres_evol_flag;

    2	2	    1348
    1	NULL	184
    1	1	    2431
    0	NULL	465
    0	0	    1162
    0	1	    1116
\end{comment}

\begin{figure}
	\centering
       
        \includegraphics[width=0.45\textwidth]{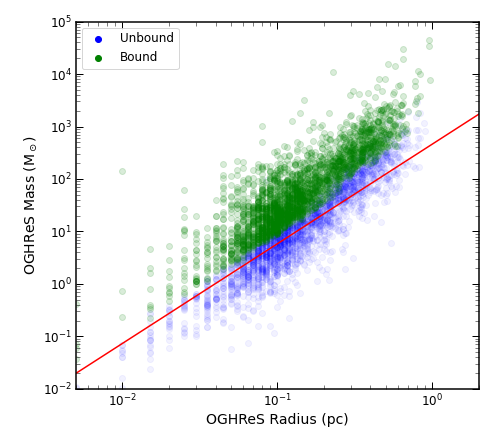}
            
    	\caption{Distinguishing between bound and unbound \higal\ clumps located the \oghres\ region. The red line shows Larson's third law (i.e. $M(r) = 460$\,\msun\,$(r/{\rm pc})^{1.9}$), which is used to distinguish between bound and unbound clumps; clumps above this threshold are considered bound while clumps below it are considered unbound. The clumps classified by \higal\ as bound and unbound are shown as green and blue circles \citep{elia2021}.} 
		\label{fig:larson_3rd_law}

\end{figure}

The modification of the physical properties has no impact on the classification of the protostellar clumps as these are identified by the presence of a 70-\mum\ point source towards their centres. However, it does affect the classification of bound and unbound starless clumps. This is mainly due to our use of the varying gas-to-dust ratio that can result in the mass of some clumps increasing significantly and changing their classification from unbound to bound. In Figure\,\ref{fig:larson_3rd_law}, we show how the changes to the physical parameters impact the \higal\ classification of clumps between unbound and bound. It is clear from this figure that a significant number of unbound clumps now have sufficient mass to be classified as bound and have been updated accordingly. These changes resulted in 1581 unbound clumps being reclassified (1116 to bound and 465 to unclassified as a distance is not available), and 184 bound clumps being unclassified as a distance is not available. This reduces the unbound population by a factor of 2. The large number of clumps being moved to the unclassified category is because in the \higal\ catalogue a distance of 1\,kpc was used to determine whether a cloud is bound or unbound when a distance had not been determined, however, we prefer not to classify clumps for which we have not been able to determined a distance to avoid any potential bias. As previously mentioned, the number of protostellar clumps remains unchanged at 1348.

\subsection{Nature of the CO non-detections}

We have not detected CO emission towards 422 clumps, which corresponds to 6.3\,per\,cent of the \higal\ sources in the OGHReS region. These are fairly evenly distributed over the whole survey longitude region. Although this is a small fraction of the full sample of \higal\ sources it is important to understand the nature of these sources to satisfy our curiosity and to provide confidence in the work presented here.

The lack of a CO detection would suggest these objects are either extragalactic in origin,  are associated with warm diffuse dust associated with evolved stars and planetary nebulae, or are low-surface density diffuse sources. The molecular line emission from extragalactic sources will be weak and is unlikely to be at a similar velocity to emission from Galactic gas. The shell of warm dust surrounding evolved stars and planetary nebula, while bright at mid- and far-infrared wavelengths, consists of low quantities of molecular gas that would fall below the detection threshold for all but the nearest evolved stars (\citealt{loup1993}). 

The extragalactic sources and evolved stars are likely to be associated with 70-\mum\ emission and be confused with protostellar objects, while the diffuse clumps will be more likely to be classified as unbound in the \higal\ catalogue. We should therefore expect the distribution of evolutionary types associated with these CO non-detections to be rather bimodal. Looking at the \higal\ evolutionary types we find 260 are unbound, 22 are bound and 140 are protostellar.

To determine the nature of the CO non-detections, we searched the SIMBAD database\footnote{http://simbad.cds.unistra.fr/simbad/} for previously classified counterparts. We found matches within a 10-arcsec radius for 56 sources, 44 of which are classified as being protostellar clumps in the \higal\ catalogue. Of these: 4 of the matches are classified in SIMBAD as active galactic nuclei (AGN) candidates and 21 more as galaxies; 9 are classified as evolved stars (i.e., mira star (1), planetary nebula (3), post-AGB star (1), carbon star (3) or emission-line star (1)); 2 are classified as young stellar objects; with the remaining eight being classified as infrared or radio sources. 
Twelve of the unbound CO non-detections are matched with a SIMBAD source (7 galaxies, 2 AGN candidates, 1 star, one emission-line star, and one YSO). 

Of the 56 matched sources, 48 have a known astronomical source within the match criterion (i.e., not just being associated with a detection at a particular wavelength), 45 have been classified as evolved stars or extragalactic in origin (corresponding to $\sim$95\,per\,cent of the matches sample). 

%select higal_evol_flag, avg(dust_temperature), stddev(dust_temperature), count(*) from oghres_higal_complete where oghres_vlsr_final > -9990 group by higal_evol_flag

\begin{figure}
    \centering
     \includegraphics[width = 0.45\textwidth]{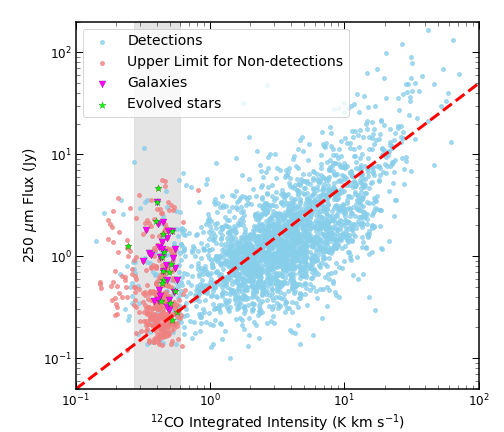}
    \caption{Comparison of the \higal\ 250-\mum\ flux and integrated $^{12}$CO (1--0) intensity. Upper limits for CO non-detections are determined assuming a flux of 3$\sigma$ per channel  extending over 3 channels. The grey shaded region indicates the mean 3--5$\sigma$ $^{12}$CO noise for the sample. Matches between the \higal\ clumps not detected in CO and external galaxies and evolved stars are also shown.}
    \label{fig:flux_flux_plot}
\end{figure}

In Figure\,\ref{fig:flux_flux_plot}, we show the correlation between the \higal\ 250-\mum\ flux and integrated $^{12}$CO intensity. Both of these fluxes arise from the dense molecular gas regions and so should be correlated and indeed this is seen in the plot and confirmed by the Spearman correlation coefficient  ($r_s = 0.58$ and $p$-value\,$ \ll 0.001$). We have performed a linear least-squares fit to the log values and included this in the plot (the slope and intercept are $0.68\pm0.02$ and $-0.18\pm0.01$, respectively). Fig.\,\ref{fig:flux_flux_plot} also includes upper limits for the CO non-detection. The distribution of the non-detections consists of a tightly concentrated knot of clumps, the location of which is consistent with the correlation found for CO detections, and a more loosely concentrated group of sources that extends to higher 250-\mum\ flux values and are significantly offset from the correlation found for CO detections.

The knot primarily consists of bound/unbound clumps while the more loosely distributed group primarily consists of protostellar clumps. The 250-\mum\ fluxes for the clumps associated with this knot are consistent with corresponding $^{12}$CO fluxes that fall below the OGHReS detection threshold and it is therefore not surprising that they have not been detected in \oghres; the majority of unbound clumps are found to be associated with this knot. The loosely distributed group is where the vast majority of matches with previously identified evolved stars and extragalactic sources have been found. These sources are characterised by higher 250-\mum\ fluxes and the presence of 70-\mum\ point sources.  If the emission detected towards these sources does originate from a circumstellar shell of warmer dust or from galaxies, as has been found for $\sim25$\,per\,cent of sources in this group, we might also expect the mean temperature to be higher than for the embedded protostellar sources; this is indeed the case with a mean dust temperature of $21.1\pm4.5$\,K and $15.2\pm4.4$\,K, respectively. Although this does not conclusively prove that all of the \higal\ protostellar clumps not detected in CO are evolved stars or extragalactic objects, it does provide strong support for this explanation.

It is worth noting that although the CO non-detections are only 6.3\,per\,cent of the whole sample, they contribute a slightly higher fraction of the unbound and protostellar clump populations, 9.7\,per\,cent and 10.4\,per\,cent, respectively. If the sample considered here is representative of the whole \higal\ catalogue these results provide an estimate of the level of contamination from evolved stars and extragalactic objects and the impact of sensitivity limits in detecting low-density clumps. We note that the contamination from extragalactic sources is likely to be lower towards the inner disc due to higher levels of interstellar extinction and so the contamination of the protostellar clump population from extragalactic background sources is expected to be lower.

\begin{comment}
0	260	0.01185962
2	140	0.11555929
1	22	0.09938182

select  higal_evol_flag, count(*),avg(SURF_DENS_NEW) from oghres_higal_complete t1,higal_clump_cat_elia18_combined t2  where t1.higal_name = t2.higal_name and oghres_vlsr_final > -99 group by higal_evol_flag

0	2423	0.02
2	1205	0.26785137
1	2565	0.19759766

There are 1205+140 = 1345

This means there is potentially 140/1345 = 10.4% contamination in the HIGAL catalogue from evolved stars and extragal objects.

\end{comment}

\begin{table*}
\centering
\caption{Median values for physical parameters for the high-reliability \higal\ catalogue \citep{elia2021}, subdivided by evolutionary class and inner/outer Galaxy location. This is an updated version of Table\,3 presented in Paper\,I where the outer Galaxy values determined from clumps with Galactocentric and heliocentric distances constrained within a factor of two. The inner Galaxy parameters are unchanged.}
\label{tbl:inner_vs_outer}
\begin{tabular}{lccc c ccc}
\hline
 & \multicolumn{3}{c}{Inner Galaxy} & &\multicolumn{3}{c}{Outer Galaxy} \\
\cline{2-4} \cline{6-8}
 &\multicolumn{1}{c}{Unbound}& \multicolumn{1}{c}{Bound} &\multicolumn{1}{c}{Protostellar} & &  \multicolumn{1}{c}{Unbound} & \multicolumn{1}{c}{Bound} & \multicolumn{1}{c}{Protostellar} \\
\hline
 \multicolumn{8}{c}{All} \\
 \hline 
$M_{\rm clump}\, [{\rm M}_\odot] $ & 27.7 & 226 & 218 &&  6.23 & 78.4 & 107 \\
$L_\mathrm{bol} [{\rm L_\odot}]$ & 33.5 & 56.3 & 553 && 2.23 & 5.87 & 86.3 \\
$L_\mathrm{bol}/M_{\rm clump}\, [{\rm L_\odot/M_\odot}]$ & 1.17 & 0.25 & 2.76 && 0.40 & 0.06 & 0.99 \\
$\Sigma$\, [g\,cm$^{-2}$]  & 0.03 & 0.14 & 0.22 && 0.03 & 0.14 & 0.29 \\
\hline 
 \multicolumn{8}{c}{ $D_{\rm heliocentric} < 8.15$\,kpc}\\
\hline 
$M_{\rm clump}\, [{\rm M}_\odot] $ & 16.7 & 109 & 94.6 && 6.14 & 66.4 & 87.0 \\

$L_\mathrm{bol} [{\rm L_\odot}]$ & 20.5 & 24.9 & 189 && 2.20 & 4.78 & 75.3 \\

$L_\mathrm{bol}/M_{\rm clump}\, [{\rm L_\odot/M_\odot}]$ & 1.05 & 0.22 & 2.46 && 0.39 & 0.06 & 1.01 \\
$\Sigma$\, [g\,cm$^{-2}$]  & 0.03 & 0.15 & 0.26 && 0.03 & 0.15 & 0.28 \\
\hline
 \multicolumn{8}{c}{ $5\,{\rm kpc} <D_{\rm heliocentric} < 7\,{\rm kpc}$}\\
\hline 
$M_{\rm clump}\, [{\rm M}_\odot] $ & 55.9 & 279 & 264 &&  58.4 & 305 & 209 \\

$L_\mathrm{bol} [{\rm L_\odot}]$ & 47.6 & 66.2 & 561 && 22.7 & 18.7 & 225 \\

$L_\mathrm{bol}/M_{\rm clump}\, [{\rm L_\odot/M_\odot}]$ & 0.80 & 0.23 & 2.53 && 0.40 & 0.06 & 1.34 \\

$\Sigma$\, [g\,cm$^{-2}$]  & 0.03 & 0.13 & 0.22 && 0.03 & 0.16 & 0.21 \\
\hline
\end{tabular}

\end{table*}

\begin{figure*}
    \centering
    \includegraphics[width = 0.45\textwidth, trim=0 0 0 0]{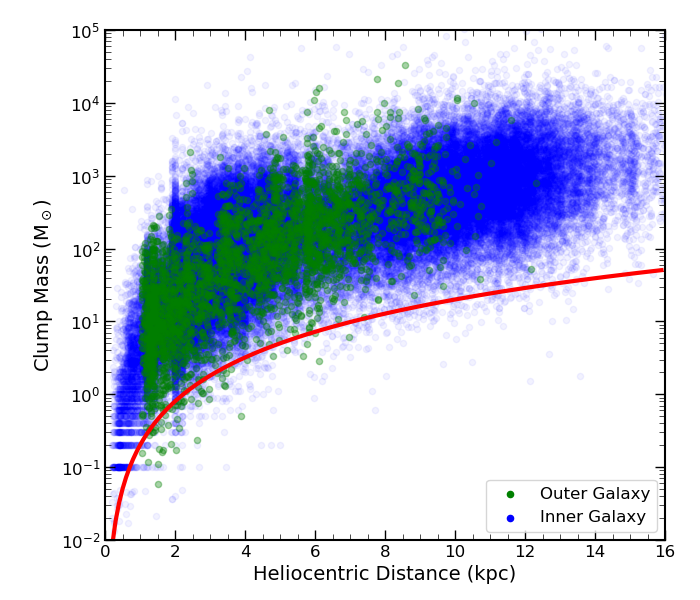}
    \includegraphics[width = 0.45\textwidth, trim=0 0 0 0]{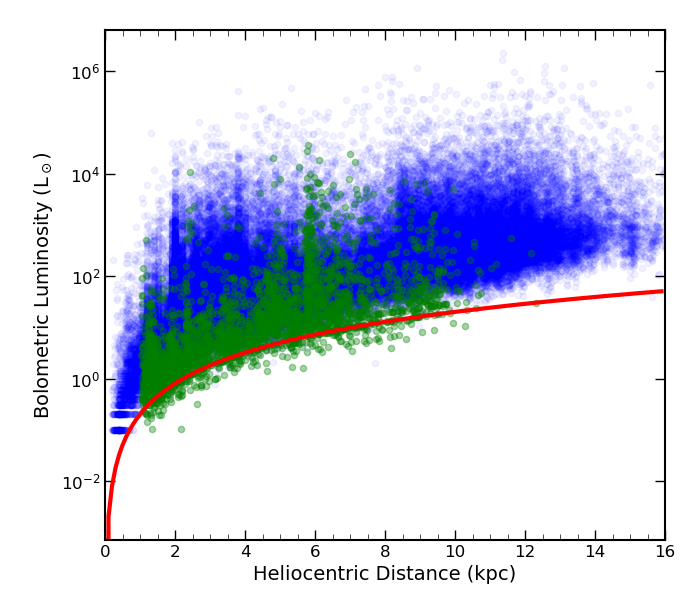}
    \includegraphics[width = 0.45\textwidth, trim=0 0 0 0]{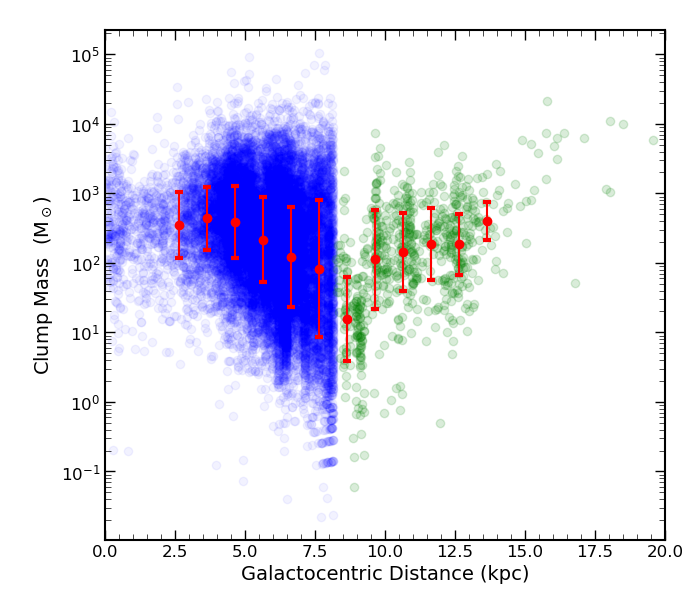}
    \includegraphics[width = 0.45\textwidth, trim=0 0 0 0]{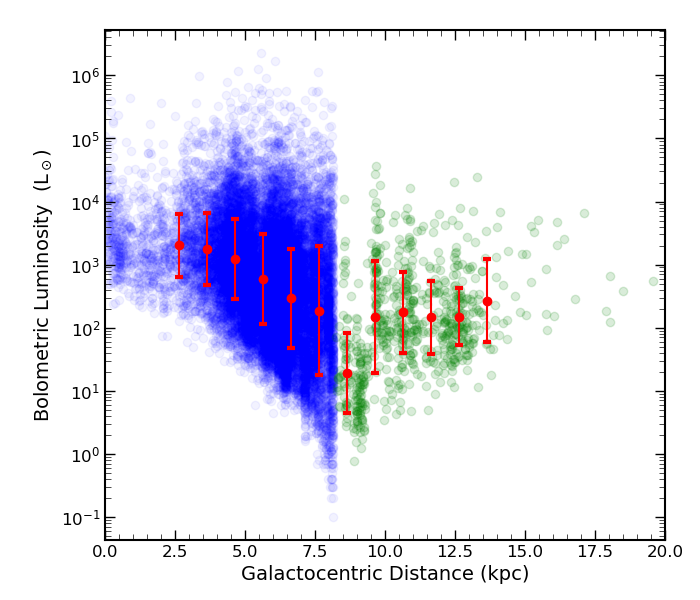}
    \caption{Physical parameters of \higal\ protostellar clumps as a function of heliocentric and Galactocentric distance. These have been updated with a gas-to-dust ratio relationship determined by \citet{giannetti2017} and the OGHReS distances presented in this work and Paper\,I. The red curves on the upper panels illustrate the impact of \higal\ flux sensitivity on the clump mass and luminosity detection limit as a function of distance (values are arbitrary). The red circles and error bars shown in the lower panels are the lognormal mean and standard deviation calculated from the clumps within each 1\,kpc wide bins between 2 and 20\,kpc. 
    }
    \label{fig:parameters_fn_rgc}
\end{figure*}

\section{Discussion}
\label{sect:discussion}

\subsection{Galactic trends revisited}

\begin{figure}
    \centering
    
     \includegraphics[width = 0.42\textwidth, trim=0 0 0 0]{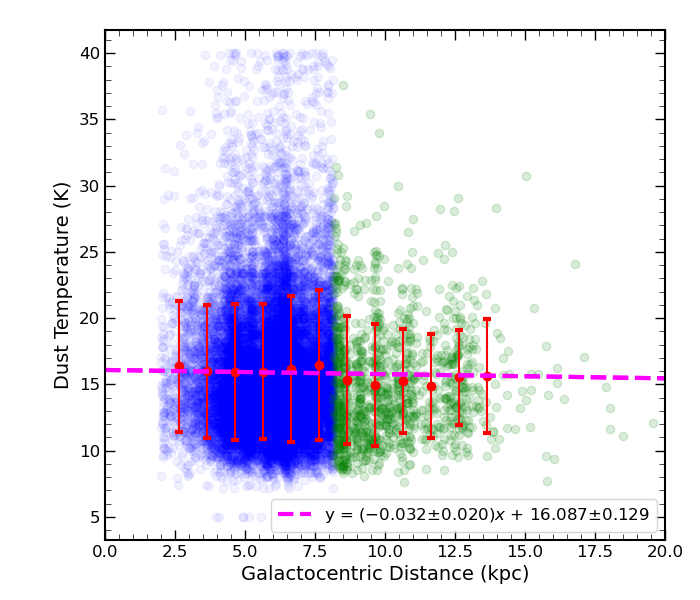}
    \includegraphics[width = 0.42\textwidth, trim=0 0 0 0]{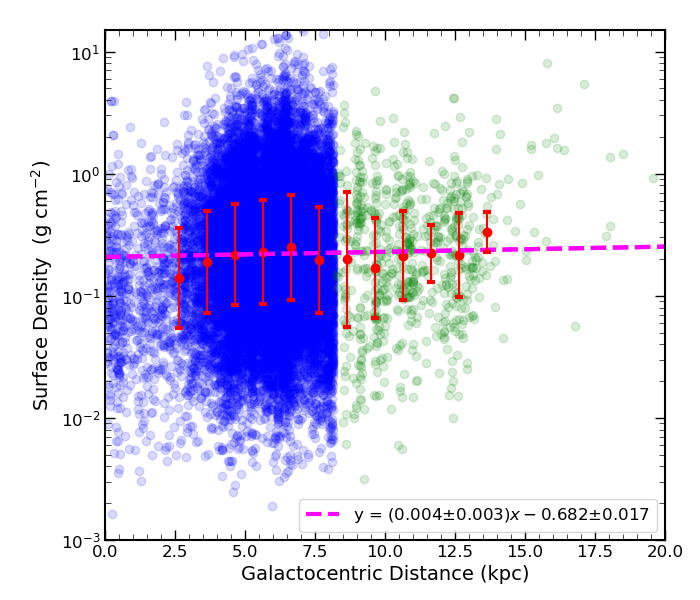}
    \includegraphics[width = 0.42\textwidth, trim=0 0 0 0]{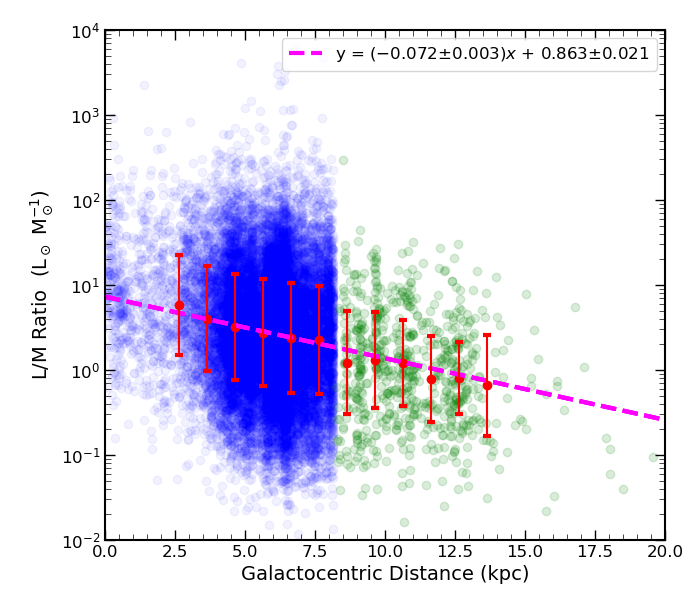}
    \caption{Distance independent parameters of \higal\ protostellar clumps as a function of Galactocentric distance (blue and green circles indicate properties taken from the \higal\ catalogue and determined here, respectively). These have been updated with a gas-to-dust ratio relationship determined by \citet{giannetti2017} and the OGHReS distances presented in this work and Paper\,I. The red circles and error bars shown in all three panels show the mean values and standard deviation calculated from the clumps within each 1-kpc wide bin between 2.5 and 13.5\,kpc (in the middle and lower panels these are determined from lognormal values). The dashed magenta lines show the results of linear least-squares fits to the plotted data and the fit parameters are given in the respective insets. The fits have been made to all data with Galactocentric distances larger than 3\,kpc and Galactic longitudes between 5\degr\ and 355\degr\ in order to exclude sources towards the Galactic centre region where distances are not considered to be reliable.
    }
    \label{fig:dist_independent_parameters_fn_rgc}
\end{figure}

The improvements to the velocity assignments, both in terms of reliability and completeness, resulted in a significant increase in the robustness of the physical properties of the \higal\ clumps in this part of the Galaxy. In Paper\,I, we presented a detailed discussion of how these improvements change our interpretation of the properties of the inner and outer Galactic populations of dense clumps. Some of the key findings were that the masses of the inner and outer Galaxy clumps are very similar, while the luminosities of inner Galaxy clumps are a factor of two to three times higher than in the outer Galaxy. Furthermore, the surface density for protostellar clumps was invariant across the Galactic disc (cf. \citealt{benedettini2021}), but the star formation efficiency decreases with distance from the Galactic centre (\citealt{ragan2016}), with clumps in the outer Galaxy having a luminosity-to-mass ratio a factor of two lower than found in the inner Galaxy (cf. \citealt{djordjevic2019, brand1991}).  These conclusions were drawn from the inner Galaxy population of \higal\ clumps and the updated parameters of \higal\ clumps with $\ell$ between 250\degr\ and 280\degr.

We have repeated this comparative analysis for all \higal\ clumps in the whole \oghres\ survey region, which includes approximately twice as many clumps and extends the Galactocentric coverage to $\sim$20\,kpc. Updated plots of the heliocentric and Galactocentric distribution of the clump masses and luminosities for the protostellar clumps are shown in Fig.\,\ref{fig:parameters_fn_rgc}. All of these plots show the same overall trends reported in Paper\,I, revealing that the properties of clumps distributed throughout the third quadrant are similar. We note that both the luminosity and mass distributions have minima at $R_{\rm gc} \sim 8.15$\,kpc, this is due to the ability of the observations to discern more readily low and luminosity mass clouds locally. To avoid this bias affecting our results we define a distance-limited sample (i.e. $5\,{\rm kpc} <D_{\rm heliocentric} < 7\,{\rm kpc}$). In Table\,\ref{tbl:inner_vs_outer}, we present the inner and outer Galaxy median values for clump mass, luminosity, surface density, and luminosity-to-mass ratio for the three different clump evolutionary stages defined by \citet{elia2021}. These are separated into three different populations: all clumps are included in the upper third, all clumps located within 8.15\,kpc are given in the middle third, and in the lower third we give the values for the distance-limited sample.  

Although the values for the enlarged outer Galaxy sample are a little different to those determined in Paper\,I, they are broadly similar, the masses and luminosities are a little higher, but this is a result of excluding clumps with large distance uncertainties, which was not done in the previous paper, and tends to predominantly affect nearby clumps. Given the dependence of the masses and luminosities on distance and the relative sparseness of the outer Galaxy population with respect to the inner Galaxy population (see the upper panels of Fig.\,\ref{fig:parameters_fn_rgc}) the distance-limited sample is the most appropriate comparison of the two populations, although the values themselves are somewhat arbitrary. Using this to compare the properties of the inner and outer Galaxy clump populations we find that the masses are consistent with the inner population being a little more massive, and the luminosity of inner Galaxy clumps being a factor of 2--3 more luminous. We note that the difference in the gas-to-dust ratio between the inner and outer Galaxy means that \higal\ is less sensitive to mass in the outer Galaxy, which biases the clumps masses in the outer Galaxy to higher values. 

It is interesting to note that the median mass of the unbound clumps is significantly lower than those of sources in the other two evolutionary stages, which are broadly similar to each other. The unbound clumps have a similar luminosity to bound clumps, however, their significantly lower clump masses result in them having a significantly higher luminosity-to-mass ratio (4-7 times larger). The $L/M$-ratio is tightly correlated to dust temperature (\citealt{urquhart2018}) so these clumps are warmer than the bound clumps and their luminosity is likely to be dominated by thermal emission resulting from heating by the interstellar radiation field, which can penetrate more deeply than for bound clumps. This reinforces the distinction between bound and unbound clump populations.

In Figure\,\ref{fig:dist_independent_parameters_fn_rgc}, we show plots of the distance-independent parameters as a function of Galactocentric distance for the protostellar clump population.  The temperature distribution was previously reported by \citet{elia2021} and is unchanged, however, the surface density and $L/M$-ratio are significantly different. These plots show the dust temperature and surface densities are invariant across both the inner and outer Galactic disc. We note a slight increase in the slope of the surface density with Galactocentric distance, however, within the uncertainty, this is consistent with a slope of zero and so is not considered significant (slope $= 0.008\pm 0.003$). In contrast to the surface density and dust temperature, the $L/M$-ratio shows a steadily decreasing trend over the whole disc. This is comparable with the decrease in the star formation rate reported by \citet{elia2022} between 4\,kpc and 16\,kpc. 

The $L/M$-ratio often used as a proxy for the instantaneous star-formation efficiency (\citealt{molinari2008}, \citealt{urquhart2013_methanol}), it is effectively a measure of the efficiency a clump is turning dense material into stars. A lower value might indicate that the star formation is slower and that the fraction of clumps hosting protostars is lower in the outer Galaxy, however, in Fig.\,\ref{fig:dist_independent_parameters_fn_rgc} we are only considering protostellar clumps and so the difference must be related to the types of stars being formed. High-mass stars provide a disproportionate fraction of the luminosity in a protocluster and the lower $L/M$-ratio might indicate the presence of a few high-mass star-forming clumps in the outer Galaxy. The lack of high-mass star formation in the outer disc was reported by \citet{brand1991} from analysis of the infrared properties of molecular clouds. This early work has been supported by the low numbers of methanol masers (\citealt{green2012_mmb}) and \hii\ regions (\citealt{urquhart_radio_south, urquhart_radio_north}) being found in the outer Galaxy. More recently, \citet{djordjevic2019} reported a similar decreasing trend in the $L/M$-ratio for both bolometric and Lyman luminosities as a function of Galactocentric distance. The decrease in the Lyman photon flux confirms there is less high-mass star formation per unit volume taking place in the outer Galaxy.

\subsection{Star Formation Fraction}

The star formation fraction (SFF) is the fraction of dense clumps associated with protostellar objects per unit volume (\citealt{ragan2016}). This is a distance-independent ratio that is related to the instantaneous star formation efficiency (SFE) that can be used to evaluate how the star formation varies as a function of Galactocentric radius. \citet{ragan2016} use the ratio of \higal\ protostellar clumps to the full population of \higal\ clumps in circular bins centred on the Galactic centre to calculate the SFF as a function of Galactocentric radius. That work was based on an earlier version of the \higal\ compact source catalogue that covered the inner Galaxy ($14\degr < \ell < 67\degr$ and $293\degr < \ell < 350\degr$; \citealt{molinari2016}).

\citet{ragan2016} found a linear gradient in the SFF within the inner Galaxy, decreasing from approximately 0.3 at a Galactocentric distance of 3\,kpc to around 0.15 at 8.5\,kpc. This corresponds to a gradient of $-0.026\pm0.002$\,kpc$^{-1}$. The lack of any significant variation at the tangent distances of spiral arms led them to suggest that spiral arms have little effect on the star formation in dense clumps. \citet{elia2021} use the full catalogue to explore the SFF out to larger Galactocentric distances (see upper-left panel of their Figure\,28) and found it increases sharply outside of the \SC\ with a peak at $\sim$11\,kpc. Their analysis includes clumps seen in the outer Galactic disc on the far side of the Galactic centre, which are likely to include a larger fraction of protostellar objects due to detection limits. We have repeated their analysis concentrating on the clumps in the  \oghres\ region using the velocity and distance refinements described earlier.

\begin{figure}
    \centering
   \includegraphics[width = 0.45\textwidth, trim=0 0 0 0]{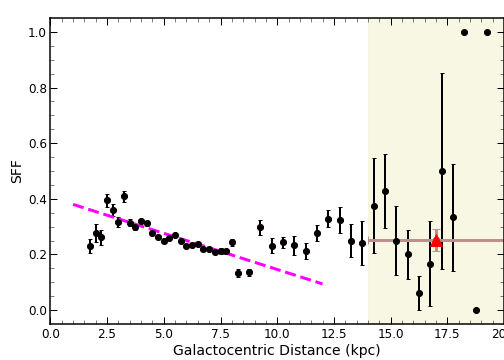}

    \caption{Star formation fraction as a function of Galactocentric radius. The dashed magenta line shows the fit reported by \citet{ragan2016} for the 3--8.5\,kpc range. The region highlighted in beige indicates where the uncertainties are too large and the data have been excluded from the analysis. The red triangle shows the SFF determined from the 14\,kpc to 20\,kpc region to provide an indication of the direction of travel in this region of the far outer Galaxy. The uncertainties are determined from binomial statistics and for bins with low numbers of sources of the same type these tend to zero.
    }
    \label{fig:sff_rgc}
\end{figure}

In Figure\,\ref{fig:sff_rgc}, we show the SFF as a function of Galactocentric distance. For the inner part of the disc (i.e., 2--8.25\,kpc), we used the \citet{elia2021} \higal\ catalogue, employing the same longitude ranges as the original \citet{ragan2016} study. This resulted in a slightly larger sample size (67\,479 sources), but the correlation between this modified sample and the fit reported by \citet[][dashed magenta line]{ragan2016} is robust. For Galactocentric distances beyond 8.25\,kpc, SFF values were derived from the \oghres\ clumps. Given the substantial uncertainties beyond approximately 14\,kpc, our analysis is limited to the region between 8.25\,kpc and 14\,kpc. While the fitted slope accurately represents the SFF between 3 and 8\,kpc, a significant deviation occurs beyond 8\,kpc, with elevated SFF values and peaks at approximately 9.5\,kpc and 12.5\,kpc. These peaks correspond to local maxima observed in the source count distribution (see the lower panel of Figure\,\ref{fig:distance_velocity_hist}), with the 9.5-kpc peak associated with the Local arm and the latter peak encompassing clumps from both the Outer and Perseus arms. Although we do not discuss the SFF for the bins beyond 14\,kpc due to large associated uncertainties we do include a point (red triangle) that shows the SFF for all sources between 14\,kpc and 20\,kpc; this point has a SFF of 0.25 and is similar to the average value between 8\,kpc and 14\,kpc, indicating the SFF is similar across the outer Galactic disc.

The observed increase in the SFF in the outer Galaxy aligns broadly with the findings of \citet{elia2021}. Differences in the number and location of SFF distribution peaks beyond 8\,kpc can probably be attributable to variations in the sampled regions. This rise in the outer disc is unexpected, given the consistent decrease observed in the inner disc, and lacks a straightforward explanation. \citet{ragan2016,ragan2018} rigorously investigated potential biases that could influence the SFF trend in the inner Galaxy but found no factors that could account for it. While they identified some enhancements in the SFF near spiral arm tangents, these were attributed to localized massive star formation complexes, leading to the conclusion that spiral arms do not enhance star formation (\citealt{ragan2018}). Similarly, the SFF peaks observed in the outer Galaxy may be associated with such complexes rather than spiral arm structures, although a detailed investigation is beyond the scope of the current work and will be revisited in a future paper.

The contrasting SFF trends in the inner and outer Galaxy present an intriguing puzzle in the context of star formation. The decreasing gradient in the inner disc, as reported by \citet{ragan2016}, is particularly perplexing as one might expect the opposite trend based on large-scale environmental variations. Factors such as decreased metallicity, a weaker interstellar radiation field, lower shear forces, and reduced thermal and turbulent pressure at larger Galactocentric radii would generally be anticipated to promote, rather than hinder, star formation, as discussed by \citet{ragan2016,ragan2018}. Furthermore, the prevailing view suggests that star formation within dense molecular clumps proceeds largely independently of the broader galactic environment (e.g., \citealt{urquhart2018}).

A potential explanation for the increasing SFF in the outer Galaxy lies in the variation of the relative durations of the quiescent and active stages of cluster formation. If star clusters forming within molecular clumps in the outer Galaxy spend a more extended period embedded within their natal material, the observed SFF would naturally increase. This is because a longer embedded phase increases the likelihood of detecting sources exhibiting both submillimetre continuum emission (indicating cold dust) and 70~$\mu$m emission (tracing warmer dust heated by more evolved young stellar objects). A key difference between the inner and outer Galaxy that could drive this variation is the significantly lower abundance of high-mass stars in the outer parts of the disc.

High-mass stars are recognized as a crucial source of feedback, capable of dispersing and dissociating the molecular gas of their parent clumps, or at least significantly accelerating this process. The relative scarcity of high-mass star formation sites in the outer Galaxy, as evidenced by studies like \citet{wouterloot1995} and \citet{urquhart2014_rms}, suggests that clump dispersal may proceed more slowly in these regions. This slower dispersal could result in longer embedded stage and, consequently, the higher SFF observed. While highly speculative, this idea of a prolonged embedded stage in the outer Galaxy is supported by observational studies by \citet[][ATLASGAL]{wells2022} and \citet[][\higal]{veneziani2017}, and theoretical work by \citet{dib2013}, all of whom found that lower-mass clumps tend to have higher star formation efficiencies. Given that high-mass stars form in the most massive clumps (e.g., \citealt{urquhart2013_cornish, urquhart2014_atlas}), their lower SFE is consistent with the higher levels of feedback associated with the presence of high-mass stars \citep{wells2022}. Thus, in the absence of strong feedback from the most massive stars, the star formation process may proceed for an extended period in the outer Galaxy. Determining the ages of embedded objects across a large sample of clumps in both the inner and outer Galaxy could potentially reveal a statistically significant difference, with objects in the outer Galaxy exhibiting older average ages and possibly a wider age distribution. 

This proposed mechanism, which hinges on the differing high-mass star content across the Galaxy, appears to contradict the notion that star formation within clumps is largely independent of the environment. However, the feedback from massive stars can be considered a local effect, internal to each clump. Even if a clump is effectively decoupled from the larger-scale galactic environment, it is still inevitably influenced by the radiation, winds, and eventual supernovae of the massive stars forming within it \citep[e.g.][]{Barnes2021,Olivier2021}. Therefore, the presence or absence of massive stars represents a local, internal driver of clump evolution rather than, or in addition to, a large-scale environmental influence. 

Beyond the difference in high-mass star formation, other intrinsic properties of molecular clouds in the outer Galaxy may contribute to the observed SFF trend. These clouds are typically characterized by, e.g., lower temperatures and densities. Notably, the lower density of molecular gas in the outer Galaxy could also slow down the overall evolutionary timescale of the clouds. Assuming a simplified free-fall collapse scenario, a significant portion of the collapse time is spent in the early, low-density phase. This early stage of collapse might be particularly challenging to detect in the outer Galaxy due to the lower masses of the clouds, dust content and dust temperatures.

In summary, what is clear is that the outer Galaxy exhibits a marked increase in the proportion of clumps associated with star formation activity. At the same time, the $L/M$-ratio decreases in this region, which, as discussed earlier, may be linked to reduced high-mass star formation. Considering together the SFF, the $L/M$-ratio, and the lower masses of clouds in the outer Galaxy suggests a higher SFE in the outer Galaxy, but with the formation of lower-mass stars. 

On the other hand, the underlying reasons for the increasing SFF in the outer Galaxy, as well as those related to its decreasing trend in the inner disc \citep{ragan2016}, remain unclear. The global physical conditions differ significantly between the inner and outer Galaxy (lower metallicity, weaker interstellar radiation field, lower pressure), and we note that the change in SFF occurs near the co-rotation radius. However, understanding what causes the behaviour of the SFF across the disc remains an open question, but it is beyond the scope of the current study.

\subsection{Fraction of dense clumps located in the outer Galaxy}

\begin{figure}
	\centering

         \includegraphics[width=0.45\textwidth]{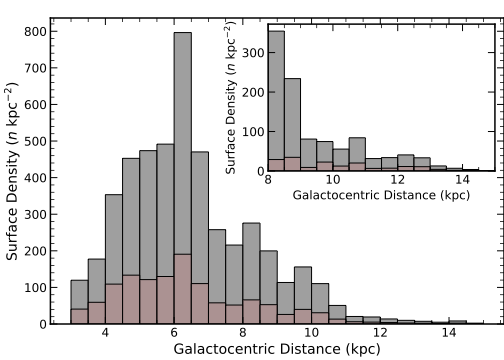}
        
    	\caption{The main plot shows the surface density distribution of clumps from the \higal\ catalogue (\citealt{elia2021}) while the insert shows the surface density distribution of clumps in the \oghres\ region. The grey histogram includes all clumps while the rose-coloured bars include only protostellar clumps. We have excluded clumps within 1.5\,kpc of the Sun to avoid contaminating the plot with a large number of nearby, low-mass clumps that would otherwise produce a strong peak at $\sim$8\,kpc. The bin width is 0.5\,kpc.}
		\label{fig:surface_density_distn}

\end{figure}

In Figure\,\ref{fig:surface_density_distn}, we show the surface-density distribution of all \higal\ clumps and those classified as being protostellar. The distribution peaks at $\sim6$\,kpc and drops off steeply with Galactocentric distance with much lower clump number surface densities in the outer Galactic disc. The number of clumps in the inner Galaxy can be taken directly from the \higal\ catalogue ($\sim$111\,000; \citealt{elia2021}). However, care must be taken when considering the number of outer Galaxy clumps, since much of the outer Galaxy is located on the far side of the disc at heliocentric distances of more than 15\,kpc. At these distances, we are only sensitive to the most massive clumps and this may lead to an underestimation of the outer Galaxy clump population and the proportion of star formation taking place in the outer Galaxy. 

If we assume that \oghres\ region is representative of the outer Galaxy, we can use the surface density of clumps in 
the surveyed area 
%this part of the Galaxy 
to estimate the total number of clumps in the outer Galaxy. We first need to determine the clump surface density by dividing the number of clumps in Galactocentric radius bins by the area of each annulus covered by \oghres.\footnote{The annuli are centred on the Galactic Centre but the area covered by the \oghres\ survey is determined by a sector of a circle centred on the position of the Sun.} The insert in Fig.\,\ref{fig:surface_density_distn} shows the outer Galaxy surface density determined from the clumps in the \oghres\ region. To estimate the number of clumps in the outer Galaxy we multiply the number of sources per radian in each annulus by $2\pi$ and then sum all of the Galactocentric radii bins. This gives $\sim$30\,000 clumps; however, due to the different latitude ranges, \oghres\ only includes approximately half of all \higal\ clumps in the longitude range 180\degr\ to 280\degr\ region and taking this into account provides a rough estimate of 60\,000 for the outer Galaxy clump population. This would suggest that 35\,per\,cent of Galactic population of clumps resides in the outer Galaxy. This is higher than a value of 25\,per\,cent that is obtained from the \higal\ catalogue but consistent with estimates of the amount of molecular gas located in the outer Galaxy ($\sim$30\,per\,cent; \citealt{heyer2015}).

The picture that is emerging is that approximately a third of the Galaxy's molecular gas and population of dense clumps is located in the outer Galaxy, and that a higher fraction of the dense clumps are associated with star formation compared to the inner Galaxy. Although the star formation activity appears to be higher in the outer Galaxy the lower $L/M$-ratio indicates that the clumps are producing lower-mass stars (see also  \citealt{brand2001}). The star formation across the disc has very different characteristics. It is unclear what is driving these observed differences, however, we know that the physical conditions in the outer and inner Galactic disc are very different and understanding this may be key to understanding the role of the environment in the star formation process.

\subsection{Potential impact on the whole Hi-GAL catalogue}

We have shown in the previous sections and in Paper\,I that using low-resolution $^{12}$CO (1--0) can produce a significant number of incorrect velocity allocations and is unable to allocate a velocity at all in many more cases. In this section we will first evaluate the impact of using low-resolution spectra in the OGHReS region and use this to estimate the potential impact on the whole \higal\ catalogue.

In the OGHReS region there are 6\,706 clumps, 956 clumps are located in the $^{12}$CO (1--0) and $^{13}$CO (1--0) Exeter FCRAO OGS region ($\ell = 180\degr$-195\degr; \citealt{mottram2010_ogs}) and 5521 are located within the overlap region with the NANTEN $^{12}$CO (1--0) data ($\ell = 205\degr$-280\degr; \citealt{mizuno2004}).

The OGS data have been used by the \higal\ team (\citealt{mege2021}) to assign velocities to 815 clumps, 736 of which are in good agreement with the \oghres\ velocities (90\,per\,cent), with 95\,per\,cent agreement for the OGS $^{13}$CO (625/649 clumps) but only 72\,per\,cent for the OGS $^{12}$CO spectra (111/154 clumps). Of the 956 clumps in the OGS region, \oghres\ has been able to allocate a velocity to 888 clumps, corresponding to 93\,per\,cent of the sample, which is only slightly better than the OGS data. Although the agreement with the OGS $^{12}$CO is poorer, velocities allocated by the \higal\ team using this transition represent a much smaller proportion of the OGS sample. However, the OGS $^{12}$CO data have been used to assign the velocities to 7\,260 clumps in the \higal\ catalogue and if the success of identifying the correct velocity component determined in the OGHReS region is representative then there may be 2--3 thousand clumps assigned with incorrect velocities.

Of the 5\,521 clumps in the NANTEN region, \higal\ allocated velocities to 4\,462 clumps with no velocity being allocated to the remaining 1\,059 clumps. Comparing these with the \oghres\ velocities we find agreement for 3\,358 clumps, corresponding to $\sim$61\,per\,cent of all clumps in the NANTEN region. The velocities allocated using NANTEN $^{12}$CO (1--0) data are therefore incorrect for approximately 20\,per\,cent of the clumps and inconclusive for a further 20\,per\,cent. For comparison, \oghres\ has been able to allocate velocities to 5\,085 of the 5\,521 clumps in the region covered by NANTEN, including 867 of the 1059 clumps the \higal\ study was unable to allocate a velocity for, corresponding to 92\,per\,cent of the sample. It is clear from the comparisons outlined above that assigning velocities using low-resolution $^{12}$CO data is far less reliable than using high-resolution $^{13}$CO data. 

The number of clumps with velocities allocated using the NANTEN data is 28\,193 (including the 4\,462 clumps discussed in the previous paragraph), which is a significant fraction of the \higal\ catalogue (corresponding to 22\,per\,cent of the catalogue with velocities assigned). Furthermore, the vast majority of these clumps (22\,246) are located in the fourth quadrant of the Galaxy (i.e. $280\degr < \ell <  359\degr$), which corresponds to 40\,per\,cent of the \higal\ catalogue with velocities assigned in this part of the Galaxy (55\,750). If the success rate for assigning the correct velocity from the NANTEN data in the third quadrant is representative then we can expect a similar proportion in the inner Galaxy. We have found that one in four velocity assignments in the OGHReS region made with the NANTEN data is incorrect. Extrapolating this to the whole NANTEN sample would suggest there are $\sim$7\,000 clumps with incorrect velocity assignments.

However, there are only two prominent spiral arms in the third quadrant and they are well separated in velocity for longitudes away from the anti-centre, while the fourth quadrant contains significant parts of the Norma, Sagittarius and Scutum-Centaurus spiral arms, the far side of the Galactic bar and the near and far 3-kpc arms (see Fig.\,\ref{fig:coverage_map}). The situation is made more complicated by the fact that the $^{12}$CO (1--0) emission has broad line, and the arms overlap in velocity, leading to the blending of CO components, and is optically thick and subject to self-absorption. All of these effects are likely to make assigning velocities more difficult in the inner Galaxy, and therefore, the estimated number of clumps with incorrect velocities is likely to be a lower limit to the true number.

\section{Summary and conclusions}
\label{sect:summary}

We have used data from the full OGHReS survey, which is a new outer-Galaxy molecular-line survey, to refine the velocities, distances and physical properties assigned to a large sample of dense clumps identified by the \higal\ survey (\citealt{molinari2010a, elia2021}) located in the $\ell =  180\degr-250\degr$  and 1\degr\ in latitude varying to account for the Galactic warp present in the third quadrant (i.e., $-1\degr < b <-0\degr$ at $\ell =180$\degr\ changing smoothly to $-2\degr < b <-1\degr$ at $\ell =280$\degr). 

We have extracted $^{12}$CO\,(2--1), $^{13}$CO\,(2--1) and  C$^{18}$O\,(2--1) spectra towards 3\,122 clumps and used these to determine the velocities of clumps along each line of sight. In cases where multiple components are detected towards a clump, making it difficult to assign a velocity unambiguously from the spectra alone, we created integrated emission maps of the different components and selected the velocity for which the CO-traced gas distribution best matches the position and morphology of the \higal\ clump. We have been able to allocate a reliable velocity to 2\,781 clumps (88\,per\,cent of the sample). This includes refined velocities for 288 clumps and the allocation of velocities to 411 clumps for which no velocity was previously available. Of the remaining clumps, 22 have multiple CO components and we have not been able to assign a velocity to, and no CO emission is detected towards 319 clumps.

We combine the results from this analysis with those reported in Paper\,I to produce a comprehensive and well-characterised sample of dense clumps. This consists of reliable velocities to 6\,193 clumps out of the 6\,706 clump in the OGHReS region, which corresponds to 92.3\,per\,cent of the sample. There are a total of 422 CO non-detections (6.3\,per\,cent) and 91 clumps with no velocity allocated (1.4\,per\,cent). This updated sample covers the whole 3rd quadrant ($180\degr < \ell < 280\degr$) and includes approximately half of all \higal\ clumps in the longitude range.\\

\noindent Our main findings are:

\begin{itemize}

    \item The correlation between the clumps and the spiral arm loci is much improved over the original \higal\ catalogue (\citealt{elia2021}) in the 3rd quadrant. We also detected a small but significant number of clumps located beyond the outer arm that appear to be correlated with the new Outer Scutum-Centaurus arm. These clumps therefore provide support for the existence of this newly discovered spiral structure (\citealt{sun2015,sun2024}).  \\

    \item We find the 422 non-detections are primarily unbound clumps with low-surface densities that fall below the OGHReS sensitivity limit (260 clumps) or are classified as protostellar (140 clumps) due to their association with a 70-$\mu$m point source. A search of SIMBAD reveals counterparts for 27 of the protostellar objects (within 10\,arcsec), 22 of which are classified as evolved stars or extragalactic in origin. From this analysis we conclude that $\sim$10\,per\,cent of the protostellar clump population may be contaminated by evolved stars and galaxies.  \\

    \item Reliable distances are available for 4\,922 clumps (uncertainties less than a factor of two) and these are used to recalculate the physical properties of the clumps. We have used this sample to confirm the trends of decreasing $L/M$ ratio with Galactocentric distance and the similarity of clump surface density of protostellar clumps across the Galactic disc. The higher values of $L/M$-ratio found in the inner Galaxy for protostellar clumps compared to protostellar clumps located in the outer Galaxy is consistent with higher-mass stars forming clumps in the inner disc. However, the invariance of surface density across the whole Galactic disc indicates that the conditions necessary for star formation are also similar. \\

    \item We have used the refined sample of outer Galaxy clumps to investigate the star formation fraction (SFF) out to 14\,kpc. This fraction is linked to the star-formation efficiency and has been found to decrease with Galactocentric distance in the inner disc (3-8\,kpc; \citealt{ragan2016}). However, beyond the Solar circle the values are seen to increase, with peaks at $\sim$9.5\,kpc and $\sim$12.5\,kpc. The larger SFF found in the outer disc indicates that, although the outer Galaxy clumps are forming lower-mass stars, there are overall more clumps forming stars in the outer Galaxy than in the inner Galaxy. It is unclear what is driving this increase in the SFF in the outer Galaxy; however, we know that the physical conditions in the outer and inner Galactic disc are very different and understanding this may be key to understanding the role of the environment in the star formation process. \\ 

    \item We have used the results to determine the reliability of the different survey data used by \higal\ to allocate velocities in this region and find the excellent agreement with $^{13}$CO\,(1-0)  data (95\,per\,cent) but significantly lower reliability for $^{12}$CO\,(1-0) (72\,per\,cent and 65\,per\,cent for high- and low-angular resolution spectra, respectively). Extrapolating these results to the rest of the \higal\ catalogue, we estimate there may be as many as 10\,000 clumps with incorrect velocities and consequently physical properties, the majority of these will be located in the 4th quadrant.

\end{itemize}

This work has demonstrated the utility of the OGHReS survey data and the improvement that its sensitivity and resolution has provided over what was previously available while also providing a much clearer idea of the structure of the Milky Way outside of the Solar circle.

\section*{Acknowledgements}

We thank the referee Michael G. Burton for their positive and timely response and for their comments and suggestions that have helped improve the clarity and readability of this work. We dedicate this work to Karl M. Menten, whose tireless mentorship, kindness, and open-hearted support profoundly shaped our scientific journeys and continues to inspire.  We are very grateful for the continuous
support provided by the APEX staff. We also extend our thanks for Davide Elia for useful discussions and clarifications for the \higal\ catalogue. JU acknowledges the support from the  Chinese Academy of Sciences (CAS) in the form of a PIFI fellowship awarded to work with Prof. Zhibo Jiang and colleagues at the Purple Mountain Observatory (MZ10007920240511600643).  ZJ acknowledges the support through the National Natural Science Foundation of China (NSFC, grant Nos. U2031202, 12373030, and 11873093). AK and MF acknowledge support from the Polish National Agency for Academic Exchange grants No. BPN/BEK/2021/1/00319/DEC/1 and BPN/BEK/2023/1/00036/DEC/01, respectively. MF acknowledges also support from the Polish National Science Centre via the grant UMO-2022/47/D/ST9/00419. ED, SN, MF and DC gratefully acknowledges the Collaborative Research Center 1601 (SFB 1601 sub-projects A1, B1 and B3) funded by the Deutsche Forschungsgemeinschaft (DFG, German Research Foundation) – 500700252. ED is a member of the International Max Planck Research School (IMPRS) for Astronomy and Astrophysics at the Universities of Bonn and Cologne. W.-J. K. was supported by DLR/Verbundforschung Astronomie und Astrophysik Grant 50 OR 2007. S.L. acknowledges support by the INAF PRIN 2019 grant ONSET. This research made use of \texttt{ASTROPY}\footnote{\href{https://www.astropy.org}{https://www.astropy.org}}, a community-developed core \texttt{PYTHON} package for Astronomy (\citealt{astropy2013,astropy2018}),  \texttt{NumPy} \citep{numpy}\footnote{\href{https://numpy.org}{https://numpy.org}}, \texttt{SciPy} \citep{scipy}\footnote{\href{https://www.scipy.org}{https://www.scipy.org}}, and \texttt{Matplotlib} \citep{matplotlib}\footnote{\href{https://matplotlib.org/}{https://matplotlib.org/}}.  This document was prepared using the Overleaf web application, which can be found at www.overleaf.com.
This research has made use of the VizieR catalogue, operated at CDS, Strasbourg, France. 

This publication is based on data acquired with the Atacama Pathfinder Experiment (APEX). APEX is a collaboration between the Max-Planck-Institut
f\"ur Radioastronomie, the European Southern Observatory, and the Onsala Space Observatory.

%%%%%%%%%%%%%%%%%%%%%%%%%%%%%%%%%%%%%%%%%%%%%%%%%%
\section*{Data Availability}

The \higal\ catalogue used for this work can be found at \href{http://vialactea.iaps.inaf.it/vialactea/public/}{http://vialactea.iaps.inaf.it/vialactea/public/} and the OGHReS catalogue used here can be found on CDS.

%------------------------------------------------------
\bibliographystyle{mnras}
\bibliography{urquhart2024}

\begin{thebibliography}{}
\makeatletter
\relax
\def\mn@urlcharsother{\let\do\@makeother \do\$\do\&\do\#\do\^\do\_\do\%\do\~}
\def\mn@doi{\begingroup\mn@urlcharsother \@ifnextchar [ {\mn@doi@} {\mn@doi@[]}}
\def\mn@doi@[#1]#2{\def\@tempa{#1}\ifx\@tempa\@empty \href {http://dx.doi.org/#2} {doi:#2}\else \href {http://dx.doi.org/#2} {#1}\fi \endgroup}
\def\mn@eprint#1#2{\mn@eprint@#1:#2::\@nil}
\def\mn@eprint@arXiv#1{\href {http://arxiv.org/abs/#1} {{\tt arXiv:#1}}}
\def\mn@eprint@dblp#1{\href {http://dblp.uni-trier.de/rec/bibtex/#1.xml} {dblp:#1}}
\def\mn@eprint@#1:#2:#3:#4\@nil{\def\@tempa {#1}\def\@tempb {#2}\def\@tempc {#3}\ifx \@tempc \@empty \let \@tempc \@tempb \let \@tempb \@tempa \fi \ifx \@tempb \@empty \def\@tempb {arXiv}\fi \@ifundefined {mn@eprint@\@tempb}{\@tempb:\@tempc}{\expandafter \expandafter \csname mn@eprint@\@tempb\endcsname \expandafter{\@tempc}}}

\bibitem[\protect\citeauthoryear{{Astropy Collaboration} et~al.,}{{Astropy Collaboration} et~al.}{2013}]{astropy2013}
{Astropy Collaboration} et~al., 2013, \mn@doi [\aap] {10.1051/0004-6361/201322068}, \href {http://adsabs.harvard.edu/abs/2013A%26A...558A..33A} {558, A33}

\bibitem[\protect\citeauthoryear{{Astropy Collaboration} et~al.,}{{Astropy Collaboration} et~al.}{2018}]{astropy2018}
{Astropy Collaboration} et~al., 2018, \mn@doi [\aj] {10.3847/1538-3881/aabc4f}, \href {https://ui.adsabs.harvard.edu/abs/2018AJ....156..123A} {156, 123}

\bibitem[\protect\citeauthoryear{{Barnes} et~al.,}{{Barnes} et~al.}{2021}]{Barnes2021}
{Barnes} A.~T.,  et~al., 2021, \mn@doi [\mnras] {10.1093/mnras/stab2958}, \href {https://ui.adsabs.harvard.edu/abs/2021MNRAS.508.5362B} {508, 5362}

\bibitem[\protect\citeauthoryear{{Barnes} et~al.,}{{Barnes} et~al.}{2023}]{barnes2023}
{Barnes} A.~T.,  et~al., 2023, \mn@doi [\apjl] {10.3847/2041-8213/aca7b9}, \href {https://ui.adsabs.harvard.edu/abs/2023ApJ...944L..22B} {944, L22}

\bibitem[\protect\citeauthoryear{{Benedettini} et~al.,}{{Benedettini} et~al.}{2021}]{benedettini2021}
{Benedettini} M.,  et~al., 2021, \mn@doi [\aap] {10.1051/0004-6361/202141433}, \href {https://ui.adsabs.harvard.edu/abs/2021A&A...654A.144B} {654, A144}

\bibitem[\protect\citeauthoryear{{Bloemen} et~al.,}{{Bloemen} et~al.}{1984}]{bloemen1984}
{Bloemen} J.~B.~G.~M.,  et~al., 1984, \aap, \href {https://ui.adsabs.harvard.edu/abs/1984A&A...135...12B} {135, 12}

\bibitem[\protect\citeauthoryear{{Brand} \& {Blitz}}{{Brand} \& {Blitz}}{1993}]{brand1993}
{Brand} J.,  {Blitz} L.,  1993, \aap, \href {http://cdsads.u-strasbg.fr/cgi-bin/nph-bib_query?bibcode=1993A%26A...275...67B&db_key=AST} {275, 67}

\bibitem[\protect\citeauthoryear{{Brand} \& {Wouterloot}}{{Brand} \& {Wouterloot}}{1991}]{brand1991}
{Brand} J.,  {Wouterloot} J.~G.~A.,  1991, in {Bloemen} H.,  ed.,  IAU Symposium Vol. 144, The Interstellar Disk-Halo Connection in Galaxies. p.~121

\bibitem[\protect\citeauthoryear{{Brand}, {Wouterloot}, {Rudolph}  \& {de Geus}}{{Brand} et~al.}{2001}]{brand2001}
{Brand} J.,  {Wouterloot} J.~G.~A.,  {Rudolph} A.~L.,   {de Geus} E.~J.,  2001, \mn@doi [\aap] {10.1051/0004-6361:20011057}, \href {https://ui.adsabs.harvard.edu/abs/2001A&A...377..644B} {377, 644}

\bibitem[\protect\citeauthoryear{{Churchwell} et~al.,}{{Churchwell} et~al.}{2006}]{churchwell2006}
{Churchwell} E.,  et~al., 2006, \mn@doi [\apj] {10.1086/507015}, \href {https://ui.adsabs.harvard.edu/abs/2006ApJ...649..759C} {649, 759}

\bibitem[\protect\citeauthoryear{{Colombo} et~al.,}{{Colombo} et~al.}{2021}]{colombo2021}
{Colombo} D.,  et~al., 2021, \mn@doi [\aap] {10.1051/0004-6361/202142182}, \href {https://ui.adsabs.harvard.edu/abs/2021A&A...655L...2C} {655, L2}

\bibitem[\protect\citeauthoryear{{Dame}, {Hartmann}  \& {Thaddeus}}{{Dame} et~al.}{2001}]{dame2001}
{Dame} T.~M.,  {Hartmann} D.,   {Thaddeus} P.,  2001, \mn@doi [\apj] {10.1086/318388}, \href {http://adsabs.harvard.edu/cgi-bin/nph-bib_query?bibcode=2001ApJ...547..792D&db_key=AST} {547, 792}

\bibitem[\protect\citeauthoryear{{Dib}, {Gutkin}, {Brandner}  \& {Basu}}{{Dib} et~al.}{2013}]{dib2013}
{Dib} S.,  {Gutkin} J.,  {Brandner} W.,   {Basu} S.,  2013, \mn@doi [\mnras] {10.1093/mnras/stt1857}, \href {https://ui.adsabs.harvard.edu/abs/2013MNRAS.436.3727D} {436, 3727}

\bibitem[\protect\citeauthoryear{{Djordjevic}, {Thompson}, {Urquhart}  \& {Forbrich}}{{Djordjevic} et~al.}{2019}]{djordjevic2019}
{Djordjevic} J.~O.,  {Thompson} M.~A.,  {Urquhart} J.~S.,   {Forbrich} J.,  2019, \mn@doi [\mnras] {10.1093/mnras/stz1262}, \href {https://ui.adsabs.harvard.edu/abs/2019MNRAS.487.1057D} {487, 1057}

\bibitem[\protect\citeauthoryear{{Duarte-Cabral} et~al.,}{{Duarte-Cabral} et~al.}{2021}]{cabral2021}
{Duarte-Cabral} A.,  et~al., 2021, \mn@doi [\mnras] {10.1093/mnras/staa2480}, \href {https://ui.adsabs.harvard.edu/abs/2021MNRAS.500.3027D} {500, 3027}

\bibitem[\protect\citeauthoryear{{Elia} et~al.,}{{Elia} et~al.}{2021}]{elia2021}
{Elia} D.,  et~al., 2021, \mn@doi [\mnras] {10.1093/mnras/stab1038}, \href {https://ui.adsabs.harvard.edu/abs/2021MNRAS.504.2742E} {504, 2742}

\bibitem[\protect\citeauthoryear{{Elia} et~al.,}{{Elia} et~al.}{2022}]{elia2022}
{Elia} D.,  et~al., 2022, \mn@doi [\apj] {10.3847/1538-4357/aca27d}, \href {https://ui.adsabs.harvard.edu/abs/2022ApJ...941..162E} {941, 162}

\bibitem[\protect\citeauthoryear{{Giannetti} et~al.,}{{Giannetti} et~al.}{2017}]{giannetti2017}
{Giannetti} A.,  et~al., 2017, \mn@doi [\aap] {10.1051/0004-6361/201731728}, \href {http://adsabs.harvard.edu/abs/2017A%26A...606L..12G} {606, L12}

\bibitem[\protect\citeauthoryear{{Green} et~al.,}{{Green} et~al.}{2012}]{green2012_mmb}
{Green} J.~A.,  et~al., 2012, \mn@doi [\mnras] {10.1111/j.1365-2966.2011.20229.x}, \href {http://adsabs.harvard.edu/abs/2012MNRAS.420.3108G} {420, 3108}

\bibitem[\protect\citeauthoryear{{G{\"u}sten}, {Nyman}, {Schilke}, {Menten}, {Cesarsky}  \& {Booth}}{{G{\"u}sten} et~al.}{2006}]{gusten2006}
{G{\"u}sten} R.,  {Nyman} L.~{\AA}.,  {Schilke} P.,  {Menten} K.,  {Cesarsky} C.,   {Booth} R.,  2006, \mn@doi [\aap] {10.1051/0004-6361:20065420}, \href {http://adsabs.harvard.edu/abs/2006A%26A...454L..13G} {454, L13}

\bibitem[\protect\citeauthoryear{Harris et~al.,}{Harris et~al.}{2020}]{numpy}
Harris C.~R.,  et~al., 2020, \mn@doi [Nature] {10.1038/s41586-020-2649-2}, 585, 357

\bibitem[\protect\citeauthoryear{{Heyer} \& {Dame}}{{Heyer} \& {Dame}}{2015}]{heyer2015}
{Heyer} M.,  {Dame} T.~M.,  2015, \mn@doi [\araa] {10.1146/annurev-astro-082214-122324}, \href {http://adsabs.harvard.edu/abs/2015ARA%26A..53..583H} {53, 583}

\bibitem[\protect\citeauthoryear{Hunter}{Hunter}{2007}]{matplotlib}
Hunter J.~D.,  2007, \mn@doi [Computing in Science & Engineering] {10.1109/MCSE.2007.55}, 9, 90

\bibitem[\protect\citeauthoryear{{Jackson} et~al.,}{{Jackson} et~al.}{2006}]{jackson2006}
{Jackson} J.~M.,  et~al., 2006, \mn@doi [\apjs] {10.1086/500091}, \href {http://adsabs.harvard.edu/abs/2006ApJS..163..145J} {163, 145}

\bibitem[\protect\citeauthoryear{{Kennicutt} \& {Evans}}{{Kennicutt} \& {Evans}}{2012}]{kennicutt2012}
{Kennicutt} R.~C.,  {Evans} N.~J.,  2012, \mn@doi [\araa] {10.1146/annurev-astro-081811-125610}, \href {http://adsabs.harvard.edu/abs/2012ARA%26A..50..531K} {50, 531}

\bibitem[\protect\citeauthoryear{{Loup}, {Forveille}, {Omont}  \& {Paul}}{{Loup} et~al.}{1993}]{loup1993}
{Loup} C.,  {Forveille} T.,  {Omont} A.,   {Paul} J.~F.,  1993, \aaps, \href {http://adsabs.harvard.edu/abs/1993A%26AS...99..291L} {99, 291}

\bibitem[\protect\citeauthoryear{{McClure-Griffiths}, {Dickey}, {Gaensler}  \& {Green}}{{McClure-Griffiths} et~al.}{2002}]{McClure-Griffiths2002}
{McClure-Griffiths} N.~M.,  {Dickey} J.~M.,  {Gaensler} B.~M.,   {Green} A.~J.,  2002, \mn@doi [\apj] {10.1086/342470}, \href {https://ui.adsabs.harvard.edu/abs/2002ApJ...578..176M} {578, 176}

\bibitem[\protect\citeauthoryear{{McClure-Griffiths}, {Dickey}, {Gaensler}, {Green}, {Haverkorn}  \& {Strasser}}{{McClure-Griffiths} et~al.}{2005}]{mcclure2005}
{McClure-Griffiths} N.~M.,  {Dickey} J.~M.,  {Gaensler} B.~M.,  {Green} A.~J.,  {Haverkorn} M.,   {Strasser} S.,  2005, \mn@doi [\apjs] {10.1086/430114}, \href {http://adsabs.harvard.edu/cgi-bin/nph-bib_query?bibcode=2005ApJS..158..178M&db_key=AST} {158, 178}

\bibitem[\protect\citeauthoryear{{M{\`e}ge} et~al.,}{{M{\`e}ge} et~al.}{2021}]{mege2021}
{M{\`e}ge} P.,  et~al., 2021, \mn@doi [\aap] {10.1051/0004-6361/202038956}, \href {https://ui.adsabs.harvard.edu/abs/2021A&A...646A..74M} {646, A74}

\bibitem[\protect\citeauthoryear{{Mizuno} \& {Fukui}}{{Mizuno} \& {Fukui}}{2004}]{mizuno2004}
{Mizuno} A.,  {Fukui} Y.,  2004, in {Clemens} D.,  {Shah} R.,   {Brainerd} T.,  eds,  Astronomical Society of the Pacific Conference Series Vol. 317, Milky Way Surveys: The Structure and Evolution of our Galaxy. p.~59

\bibitem[\protect\citeauthoryear{{Molinari}, {Pezzuto}, {Cesaroni}, {Brand}, {Faustini}  \& {Testi}}{{Molinari} et~al.}{2008}]{molinari2008}
{Molinari} S.,  {Pezzuto} S.,  {Cesaroni} R.,  {Brand} J.,  {Faustini} F.,   {Testi} L.,  2008, \mn@doi [\aap] {10.1051/0004-6361:20078661}, \href {http://cdsads.u-strasbg.fr/abs/2008A%26A...481..345M} {481, 345}

\bibitem[\protect\citeauthoryear{{Molinari} et~al.,}{{Molinari} et~al.}{2010}]{molinari2010a}
{Molinari} S.,  et~al., 2010, \mn@doi [\aap] {10.1051/0004-6361/201014659}, \href {http://adsabs.harvard.edu/abs/2010A%26A...518L.100M} {518, L100}

\bibitem[\protect\citeauthoryear{{Molinari} et~al.,}{{Molinari} et~al.}{2016}]{molinari2016}
{Molinari} S.,  et~al., 2016, \mn@doi [\aap] {10.1051/0004-6361/201526380}, \href {http://adsabs.harvard.edu/abs/2016A%26A...591A.149M} {591, A149}

\bibitem[\protect\citeauthoryear{{Mottram} \& {Brunt}}{{Mottram} \& {Brunt}}{2010}]{mottram2010_ogs}
{Mottram} J.~C.,  {Brunt} C.~M.,  2010, in {Kothes} R.,  {Landecker} T.~L.,   {Willis} A.~G.,  eds,  Astronomical Society of the Pacific Conference Series Vol. 438, The Dynamic Interstellar Medium: A Celebration of the Canadian Galactic Plane Survey. p.~98 (\mn@eprint {arXiv} {1007.3627}), \mn@doi{10.48550/arXiv.1007.3627}

\bibitem[\protect\citeauthoryear{{Nagayama}, {Omodaka}, {Handa}, {Honma}, {Kobayashi}, {Kawaguchi}  \& {Ueno}}{{Nagayama} et~al.}{2011}]{nagayama2011}
{Nagayama} T.,  {Omodaka} T.,  {Handa} T.,  {Honma} M.,  {Kobayashi} H.,  {Kawaguchi} N.,   {Ueno} Y.,  2011, \pasj, \href {http://adsabs.harvard.edu/abs/2011PASJ...63..719N} {63, 719}

\bibitem[\protect\citeauthoryear{{Olivier}, {Lopez}, {Rosen}, {Nayak}, {Reiter}, {Krumholz}  \& {Bolatto}}{{Olivier} et~al.}{2021}]{Olivier2021}
{Olivier} G.~M.,  {Lopez} L.~A.,  {Rosen} A.~L.,  {Nayak} O.,  {Reiter} M.,  {Krumholz} M.~R.,   {Bolatto} A.~D.,  2021, \mn@doi [\apj] {10.3847/1538-4357/abd24a}, \href {https://ui.adsabs.harvard.edu/abs/2021ApJ...908...68O} {908, 68}

\bibitem[\protect\citeauthoryear{{Poggio} et~al.,}{{Poggio} et~al.}{2024}]{poggio2024}
{Poggio} E.,  et~al., 2024, \mn@doi [arXiv e-prints] {10.48550/arXiv.2407.18659}, \href {https://ui.adsabs.harvard.edu/abs/2024arXiv240718659P} {p. arXiv:2407.18659}

\bibitem[\protect\citeauthoryear{{Ragan}, {Moore}, {Eden}, {Hoare}, {Elia}  \& {Molinari}}{{Ragan} et~al.}{2016}]{ragan2016}
{Ragan} S.~E.,  {Moore} T.~J.~T.,  {Eden} D.~J.,  {Hoare} M.~G.,  {Elia} D.,   {Molinari} S.,  2016, \mn@doi [\mnras] {10.1093/mnras/stw1870}, \href {http://adsabs.harvard.edu/abs/2016MNRAS.462.3123R} {462, 3123}

\bibitem[\protect\citeauthoryear{{Ragan}, {Moore}, {Eden}, {Hoare}, {Urquhart}, {Elia}  \& {Molinari}}{{Ragan} et~al.}{2018}]{ragan2018}
{Ragan} S.~E.,  {Moore} T.~J.~T.,  {Eden} D.~J.,  {Hoare} M.~G.,  {Urquhart} J.~S.,  {Elia} D.,   {Molinari} S.,  2018, \mn@doi [\mnras] {10.1093/mnras/sty1672}, \href {https://ui.adsabs.harvard.edu/abs/2018MNRAS.479.2361R} {479, 2361}

\bibitem[\protect\citeauthoryear{{Reid} et~al.,}{{Reid} et~al.}{2014}]{reid2014}
{Reid} M.~J.,  et~al., 2014, \mn@doi [\apj] {10.1088/0004-637X/783/2/130}, \href {http://adsabs.harvard.edu/abs/2014ApJ...783..130R} {783, 130}

\bibitem[\protect\citeauthoryear{{Reid} et~al.,}{{Reid} et~al.}{2019}]{reid2019}
{Reid} M.~J.,  et~al., 2019, \mn@doi [\apj] {10.3847/1538-4357/ab4a11}, \href {https://ui.adsabs.harvard.edu/abs/2019ApJ...885..131R} {885, 131}

\bibitem[\protect\citeauthoryear{{Rigby} et~al.,}{{Rigby} et~al.}{2016}]{rigby2016}
{Rigby} A.~J.,  et~al., 2016, \mn@doi [\mnras] {10.1093/mnras/stv2808}, \href {http://adsabs.harvard.edu/abs/2016MNRAS.456.2885R} {456, 2885}

\bibitem[\protect\citeauthoryear{{Rudolph}, {Simpson}, {Haas}, {Erickson}  \& {Fich}}{{Rudolph} et~al.}{1997}]{rudolph1997}
{Rudolph} A.~L.,  {Simpson} J.~P.,  {Haas} M.~R.,  {Erickson} E.~F.,   {Fich} M.,  1997, \mn@doi [\apj] {10.1086/304758}, \href {https://ui.adsabs.harvard.edu/abs/1997ApJ...489...94R} {489, 94}

\bibitem[\protect\citeauthoryear{{Russeil}, {Zavagno}, {M{\`e}ge}, {Poulin}, {Molinari}  \& {Cambresy}}{{Russeil} et~al.}{2017}]{russeil2017}
{Russeil} D.,  {Zavagno} A.,  {M{\`e}ge} P.,  {Poulin} Y.,  {Molinari} S.,   {Cambresy} L.,  2017, \mn@doi [\aap] {10.1051/0004-6361/201730540}, \href {https://ui.adsabs.harvard.edu/abs/2017A&A...601L...5R} {601, L5}

\bibitem[\protect\citeauthoryear{{Schuller} et~al.,}{{Schuller} et~al.}{2017}]{schuller2017}
{Schuller} F.,  et~al., 2017, \mn@doi [\aap] {10.1051/0004-6361/201628933}, \href {https://ui.adsabs.harvard.edu/abs/2017A&A...601A.124S} {601, A124}

\bibitem[\protect\citeauthoryear{{Schuller} et~al.,}{{Schuller} et~al.}{2021}]{schuller2021}
{Schuller} F.,  et~al., 2021, \mn@doi [\mnras] {10.1093/mnras/staa2369}, \href {https://ui.adsabs.harvard.edu/abs/2021MNRAS.500.3064S} {500, 3064}

\bibitem[\protect\citeauthoryear{{Stark} \& {Brand}}{{Stark} \& {Brand}}{1989}]{stark1989}
{Stark} A.~A.,  {Brand} J.,  1989, \mn@doi [\apj] {10.1086/167334}, \href {http://adsabs.harvard.edu/abs/1989ApJ...339..763S} {339, 763}

\bibitem[\protect\citeauthoryear{{Sun}, {Xu}, {Yang}, {Li}, {Du}, {Zhang}  \& {Zhou}}{{Sun} et~al.}{2015}]{sun2015}
{Sun} Y.,  {Xu} Y.,  {Yang} J.,  {Li} F.-C.,  {Du} X.-Y.,  {Zhang} S.-B.,   {Zhou} X.,  2015, \mn@doi [\apjl] {10.1088/2041-8205/798/2/L27}, \href {https://ui.adsabs.harvard.edu/abs/2015ApJ...798L..27S} {798, L27}

\bibitem[\protect\citeauthoryear{{Sun} et~al.,}{{Sun} et~al.}{2024}]{sun2024}
{Sun} Y.,  et~al., 2024, \mn@doi [\apjl] {10.3847/2041-8213/ad9605}, \href {https://ui.adsabs.harvard.edu/abs/2024ApJ...977L..35S} {977, L35}

\bibitem[\protect\citeauthoryear{{Urquhart}, {Busfield}, {Hoare}, {Lumsden}, {Clarke}, {Moore}, {Mottram}  \& {Oudmaijer}}{{Urquhart} et~al.}{2007}]{urquhart_radio_south}
{Urquhart} J.~S.,  {Busfield} A.~L.,  {Hoare} M.~G.,  {Lumsden} S.~L.,  {Clarke} A.~J.,  {Moore} T.~J.~T.,  {Mottram} J.~C.,   {Oudmaijer} R.~D.,  2007, \mn@doi [\aap] {10.1051/0004-6361:20065837}, \href {http://adsabs.harvard.edu/abs/2007A%26A...461...11U} {461, 11}

\bibitem[\protect\citeauthoryear{{Urquhart} et~al.,}{{Urquhart} et~al.}{2009}]{urquhart_radio_north}
{Urquhart} J.~S.,  et~al., 2009, \mn@doi [\aap] {10.1051/0004-6361/200912108}, \href {http://adsabs.harvard.edu/abs/2009A%26A...501..539U} {501, 539}

\bibitem[\protect\citeauthoryear{{Urquhart} et~al.,}{{Urquhart} et~al.}{2013a}]{urquhart2013_methanol}
{Urquhart} J.~S.,  et~al., 2013a, \mn@doi [\mnras] {10.1093/mnras/stt287}, \href {http://adsabs.harvard.edu/abs/2013MNRAS.431.1752U} {431, 1752}

\bibitem[\protect\citeauthoryear{{Urquhart} et~al.,}{{Urquhart} et~al.}{2013b}]{urquhart2013_cornish}
{Urquhart} J.~S.,  et~al., 2013b, \mn@doi [\mnras] {10.1093/mnras/stt1310}, \href {http://adsabs.harvard.edu/abs/2013MNRAS.435..400U} {435, 400}

\bibitem[\protect\citeauthoryear{{Urquhart}, {Figura}, {Moore}, {Hoare}, {Lumsden}, {Mottram}, {Thompson}  \& {Oudmaijer}}{{Urquhart} et~al.}{2014a}]{urquhart2014_rms}
{Urquhart} J.~S.,  {Figura} C.~C.,  {Moore} T.~J.~T.,  {Hoare} M.~G.,  {Lumsden} S.~L.,  {Mottram} J.~C.,  {Thompson} M.~A.,   {Oudmaijer} R.~D.,  2014a, \mn@doi [\mnras] {10.1093/mnras/stt2006}, \href {http://adsabs.harvard.edu/abs/2014MNRAS.437.1791U} {437, 1791}

\bibitem[\protect\citeauthoryear{{Urquhart} et~al.,}{{Urquhart} et~al.}{2014b}]{urquhart2014_atlas}
{Urquhart} J.~S.,  et~al., 2014b, \mn@doi [\mnras] {10.1093/mnras/stu1207}, \href {http://adsabs.harvard.edu/abs/2014MNRAS.443.1555U} {443, 1555}

\bibitem[\protect\citeauthoryear{{Urquhart} et~al.,}{{Urquhart} et~al.}{2018}]{urquhart2018}
{Urquhart} J.~S.,  et~al., 2018, \mn@doi [\mnras] {10.1093/mnras/stx2258}, \href {http://adsabs.harvard.edu/abs/2018MNRAS.473.1059U} {473, 1059}

\bibitem[\protect\citeauthoryear{{Urquhart} et~al.,}{{Urquhart} et~al.}{2021}]{urquhart2021}
{Urquhart} J.~S.,  et~al., 2021, \mn@doi [\mnras] {10.1093/mnras/staa2512}, \href {https://ui.adsabs.harvard.edu/abs/2021MNRAS.500.3050U} {500, 3050}

\bibitem[\protect\citeauthoryear{{Urquhart} et~al.,}{{Urquhart} et~al.}{2024}]{urquhart2024}
{Urquhart} J.~S.,  et~al., 2024, \mn@doi [\mnras] {10.1093/mnras/stad3983}, \href {https://ui.adsabs.harvard.edu/abs/2024MNRAS.528.4746U} {528, 4746}

\bibitem[\protect\citeauthoryear{{Veneziani} et~al.,}{{Veneziani} et~al.}{2017}]{veneziani2017}
{Veneziani} M.,  et~al., 2017, \mn@doi [\aap] {10.1051/0004-6361/201423474}, \href {https://ui.adsabs.harvard.edu/abs/2017A&A...599A...7V} {599, A7}

\bibitem[\protect\citeauthoryear{Virtanen et~al.,}{Virtanen et~al.}{2020}]{scipy}
Virtanen P.,  et~al., 2020, \mn@doi [Nature Methods] {10.1038/s41592-019-0686-2}, \href {https://rdcu.be/b08Wh} {17, 261}

\bibitem[\protect\citeauthoryear{{Watkins} et~al.,}{{Watkins} et~al.}{2023}]{watkins2023}
{Watkins} E.~J.,  et~al., 2023, \mn@doi [\aap] {10.1051/0004-6361/202346075}, \href {https://ui.adsabs.harvard.edu/abs/2023A&A...676A..67W} {676, A67}

\bibitem[\protect\citeauthoryear{{Wells}, {Urquhart}, {Moore}, {Browning}, {Ragan}, {Rigby}, {Eden}  \& {Thompson}}{{Wells} et~al.}{2022}]{wells2022}
{Wells} M.~R.~A.,  {Urquhart} J.~S.,  {Moore} T.~J.~T.,  {Browning} K.~E.,  {Ragan} S.~E.,  {Rigby} A.~J.,  {Eden} D.~J.,   {Thompson} M.~A.,  2022, \mn@doi [\mnras] {10.1093/mnras/stac2420}, \href {https://ui.adsabs.harvard.edu/abs/2022MNRAS.516.4245W} {516, 4245}

\bibitem[\protect\citeauthoryear{{Wouterloot}, {Fiegle}, {Brand}  \& {Winnewisser}}{{Wouterloot} et~al.}{1995}]{wouterloot1995}
{Wouterloot} J.~G.~A.,  {Fiegle} K.,  {Brand} J.,   {Winnewisser} G.,  1995, \aap, \href {https://ui.adsabs.harvard.edu/abs/1995A&A...301..236W} {301, 236}

\makeatother
\end{thebibliography}
%___________________________________________________________

\end{document}